\documentclass[preprint,onecolumn,nofootinbib]{revtex4}
\pdfoutput=1
\usepackage[colorlinks=true,linkcolor=blue,urlcolor=blue,filecolor=black,citecolor=red,pdfstartview=FitV,pdftitle={},pdfsubject={},pdfkeywords={},pdfpagemode=None,bookmarksopen=true]{hyperref}
\usepackage{graphicx}
\usepackage{amsmath}
\usepackage{amsfonts}
\usepackage{amssymb}
\usepackage{color}%
\usepackage{dcolumn}
\newcommand{\f}{\begin{equation}}
\newcommand{\ff}{\end{equation}}
\newcommand{\fa}{\begin{eqnarray}}
\newcommand{\ffa}{\end{eqnarray}}

\newcommand{\tabincell}[2]{\begin{tabular}{@{}#1@{}}#2\end{tabular}}
\begin{document}
\title{Dynamics of k-essence in loop quantum cosmology}
\author{Jiali Shi$^{1}$}
\author{Jian-Pin Wu$^{2}$}
\thanks{jianpinwu@yzu.edu.cn}
\affiliation{
$^1$ College of Physical Science and Technology, Bohai University, Jinzhou 121013, China\ \\
$^2$ Center for Gravitation and Cosmology, College of Physical Science and Technology, Yangzhou University, Yangzhou 225009, China}
\begin{abstract}

In this paper, we study the dynamics of k-essence in loop quantum cosmology (LQC). The study indicates that the loop quantum gravity (LQG) effect plays a key role only in the early epoch of the universe and is diluted at the later stage. The fixed points in LQC are basically consistent with that in standard Friedmann-Robertson-Walker (FRW) cosmology. For most of the attractor solutions, the stability conditions in LQC are in agreement with that for the standard FRW universe. But for some special fixed point, more tighter constraints are imposed thanks to the LQG effect.

\end{abstract}
\maketitle
\section{Introduction}

Numerous cosmological observations strongly suggest that the universe is undergoing accelerated expansion at present epoch \cite{Riess:1998cb,Perlmutter:1998np,Bennett:2003bz,Spergel:2003cb,Tegmark:2003ud,Tegmark:2003uf}. It is the most widely accepted idea that a mysterious dominant component, dark energy with negative pressure, drives this cosmic acceleration. The most popular dark energy model is quintessence. It is described by a canonical scalar field accompanying with a particular potential that results in the late time acceleration of the universe. However, there has been a growing interest in the study of alternative models characterized by a non-canonical kinetic term, and has made a great progress. This scenario was originally proposed to drive the inflation in the early universe \cite{ArmendarizPicon:1999rj,Garriga:1999vw}. And then, it was first applied to describe late time cosmic acceleration in \cite{Chiba:1999ka}. More general formalism of scalar field dark energy model was proposed in \cite{ArmendarizPicon:2000dh,ArmendarizPicon:2000ah} and we call it as k-essence.

As we all known, the scalar field dynamical dark energy models suffer from the so-called fine-tuning problem and coincidence problem. To attack these problems, we can resort to the scalar field models exhibiting scaling solutions, see \cite{Tsujikawa:2006mw,Gong:2006sp,Ferreira:1997hj,Copeland:1997et,Guo:2003rs,Guo:2003eu,Lin:2020fue} and therein (also see \cite{Copeland:2006wr} for a review). As a dynamical attractor, the scaling solution can partly alleviate these two problems. In addition, we can also study the stability conditions of the scalar field dynamical dark energy models, see for example \cite{UrenaLopez:2005zd,Carot:2002ww,Ito:2012}.

Lots of dynamical dark energy models, including k-essence, have been deeply studied in the framework of the standard cosmology. However, it is expected that in the regime of very high curvature, the general relativity (GR) breaks down and the big bang singularity emerges. A theory of quantum gravity shall provide us a natural scenario to attach this problem.
One of the candidate theories of quantum gravity is loop quantum gravity (LQG), a non-perturbative and background-independent quantum gravity theory \cite{LQGRovelli,LQGThiemann,Ashtekar:2004eh,Han:2005km}.
Based on the LQG, we can construct a symmetry-reduced cosmological model with homogeneous and isotropic spacetimes, known as loop quantum cosmology (LQC) \cite{Bojowald:2001xe,Ashtekar:2006rx,Ashtekar:2006wn,Bojowald:2006da,Ashtekar:2003hd,Ashtekar:2011ni}. The non-perturbative quantum gravity effects result in a $-\rho^2$ modification to the standard Friedmann dynamics.
The big bang singularity in the early universe can be resolved in this scenario
\cite{Bojowald:2001xe,Ashtekar:2003hd,Ashtekar:2006rx,Ashtekar:2006uz,Bojowald:2003xf,Singh:2003au,Vereshchagin:2004uc,Date:2005nn,Date:2004fj,Goswami:2005fu}.
It is very interesting to notice that even at the semi-classical level, instead of the big bang singularity, a big bounce emerges \cite{Bojowald:2005zk,Stachowiak:2006uh}. The LQG effect also results in the emergences of the super-inflationary phase \cite{Bojowald:2002nz}. The horizon problem with only a few number of e-foldings can be resolved in this landscape  \cite{Copeland:2007qt}. Further, to search for the potential observable prints, the cosmological perturbative theory in LQC is also deeply explored in \cite{Bojowald:2006tm,Bojowald:2006zb,Bojowald:2008gz,Bojowald:2007hv,Bojowald:2007cd,Wu:2010wj,Wu:2012mh,Wu:2018mhg} and the primordial power spectrum is studied in \cite{Copeland:2007qt,Hossain:2004wm,Calcagni:2006pr,Mulryne:2006cz,Zhang:2007bi,Artymowski:2008sc,Tsujikawa:2003vr,Shimano:2009tn}. In addition, the large scale effect of LQG is also found in \cite{Ding:2008tq}, which provides us the possibility to study the LQG effect on the dark energy evolution. Many scalar field dark energy models and their dynamics have been widely studied in the framework of LQC, see \cite{Wei:2007rp,Wu:2008db,Fu:2008gh,Chen:2008ca,Zonunmawia:2017ofc,Li:2010ju} and therein.

In this paper, we study the dynamics of k-essence and its attractor solutions in LQC. Our paper is organized as what follows. In Section \ref{sec-kessence}, we introduce the k-essence in the LQC framework and derive the equations of motion of the dynamical system. And then, we study the dynamics of k-essence for the constant coupling parameters and dynamical changing coupling parameters in Section \ref{sec-con} and Section \ref{sec-dynmics}, respectively. The conclusions and discussions are briefly presented in Section \ref{sec-con-dis}.

\section {K-essence in LQC}\label{sec-kessence}

In quintessence scalar field dark energy model, the potential energy of the scalar field plays a key role in driving the cosmic late-time acceleration. If we introduce a non-canonical kinetic energy term in the Lagrangian, we find that even when the potential vanishes, the cosmic acceleration can also achieved. This model characterized by a non-canonical kinetic energy term is called as k-essence. The most general Lagrangian of k-essence is a function of the scalar field $\phi$ and its kinetic energy term $X\equiv\frac{1}{2}\partial_{\mu}\phi\partial^{\mu}\phi$, i.e., $\mathcal{L}=f(\phi,X)$. In this paper, we consider a specific form of k-essence as \cite{ArmendarizPicon:2000dh,ArmendarizPicon:2000ah,Chakraborty:2019swx}
\begin{eqnarray}
\label{L-k-essence}
\mathcal{L}=\alpha(\phi)X+\beta(\phi)X^2-V(\phi)\,,
\end{eqnarray}
where $V(\phi)$ is the potential, and the coefficients $\alpha$ and $\beta$ are the functions of scalar field.
The above Lagrangian is a polynomial of degree $2$ in the kinetic energy $X$. This type of Lagrangian can emerge from the low-energy effective string theory \cite{ArmendarizPicon:1999rj,Garriga:1999vw}.

Incorporating the LQG effect, instead of the standard Friedmann equation, we have an effective one \cite{Bojowald:2001xe,Ashtekar:2006rx,Ashtekar:2006wn,Bojowald:2006da,Ashtekar:2003hd,Ashtekar:2011ni}\footnote{We would like to point out that there is an alternative version of modified Friedmann equation proposed in \cite{Ding:2008tq,Yang:2009fp}, which could be derived from full LQG as recently shown by \cite{Assanioussi:2018hee}.}
\begin{eqnarray}
\label{Fre-eom-lqc}
H^2={\frac{\rho_t}{3M_p^2}}\Big(1-{\frac{\rho_t}{\rho_{c}}}\Big)\,,
\end{eqnarray}
where $H\equiv\dot{a}/a$ is the Hubble parameter and $\rho_t$ is the total energy density of the cosmological contents. $M_p^2=\frac{1}{8\pi G}$ is the square of Plank mass. In what follows, we set $M_p^2=1$ for convenience. $\rho_{c}={\sqrt{3}}/(16{\pi}{\gamma}^3G^2{\hbar})$ is the critical density where $\gamma$ is the dimensionless Barbero-Immirzi parameter and $\hbar$ is Plank constant. Along with the conservation law
\begin{eqnarray}
\label{con-law}
\dot{\rho_t}+3H({\rho_t}+p_t)=0\,,
\end{eqnarray}
the effective Friedmann equation provides a description of the universe incorporating the LQG effect. $p_t$ in the above equation is the total pressure. Differentiating the Friedmann equation \eqref{Fre-eom-lqc} and using the conservation law \eqref{con-law}, one achieves the following effective Raychaudhuri equation
\begin{eqnarray}
\label{Fre-eom-lqc-v2}
\dot{H}=-{\frac{(\rho_t+p_t)}{2}}\Big(1-{\frac{2\rho_t}{\rho_{c}}}\Big)\,.
\end{eqnarray}

We assume that the contents of the universe include the k-essence scalar field and the dark matter. So the total energy density and pressure of the contents of the universe are
\fa
&&
\rho_t=\rho_\phi+\rho_m\,,
~~~~~~
p_t=p_\phi+p_m\,,
\ffa
where $\rho_\phi$ ($\rho_m$) and $p_\phi$ ($p_m$) are the energy density and pressure of the dark energy (dark matter), respectively. From the Lagrangian of k-essence \eqref{L-k-essence}, we can easily derive $\rho_\phi$ and $p_\phi$ in flat Friedmann-Robertson-Walker (FRW) background as \cite{ArmendarizPicon:2000dh,ArmendarizPicon:2000ah,Chakraborty:2019swx}
\begin{eqnarray}
&&
\rho_\phi={\alpha(\phi)}X+3{\beta(\phi)}X^2+V{(\phi)}\,,
\
\\
&&
p_\phi={\alpha(\phi)}X+{\beta(\phi)}X^2-V{(\phi)}\,.
\end{eqnarray}
Since we have assumed that the universe is homogeneity and isotropy, one has $X={\frac{1}{2}}{\dot{\phi}^2}$ where the dot denotes the derivative with respect to the time. Further, taking the variation of the Lagrangian \eqref{L-k-essence} with respect to the scalar field $\phi$, one obtains the Klein-Gordon (KG) equation as
\begin{eqnarray}
\label{KG}
\ddot{\phi}[{\alpha(\phi)}+3{\beta(\phi)}\dot{\phi}^2]+\alpha{'}{(\phi)}{\frac{\dot{\phi}^2}{2}}+3{\beta{'}(\phi)}{\frac{\dot{\phi}^4}{4}}+3H\dot{\phi}[{\alpha(\phi)}+{\beta(\phi)}\dot{\phi}^2]+V{'}{(\phi)}=0\,.
\end{eqnarray}

To study the dynamics of the above system, we define the following set of dimensionless variables:
\begin{eqnarray}
&&
x={\frac{\alpha(\phi)\dot{\phi}^2 }{6H^2}},~~~~y={\frac{\beta(\phi)\dot{\phi}^4 }{12H^2}}\,,
\nonumber
\\
&&
b={\frac{V(\phi)}{3H^2}},~~~~\lambda={\frac{1}{\alpha}}{\frac{d\alpha}{d\phi}}{\frac{\dot{\phi}}{H}}\,,
\nonumber
\\
&&
\delta={\frac{1}{\beta}}{\frac{d\beta}{d\phi}}{\frac{\dot{\phi}}{H}},~~~~z={\frac{\rho_t}{\rho_{c}}}\,.
\end{eqnarray}
In term of the above dimensionless variables, the effective Friedman equation \eqref{Fre-eom-lqc} and Raychaudhuri equation \eqref{Fre-eom-lqc-v2} can be rewritten as
\begin{eqnarray}
&&
\label{Fre-eom-lqc-new}
\frac{\rho_m}{3H^2}=\frac{1}{1-z}-(x+3y+b)\,,
\
\\
&&
\label{Ray-eom-lqc-new}
\frac{\dot{H}}{H^2}=-\frac{3}{2}\Big[2x+4y+\frac{1-(x+3y+b)(1-z)}{1-z}\Big](1-2z)\,.
\end{eqnarray}
And then, we recast the above system into the following autonomous form
\begin{eqnarray}
&&
\label{xpn}
x{'}=x{\lambda}+3x{\mathcal{G}}-2x{\mathcal{F}}\,,
\
\\
&&
\label{ypn}
y{'}=y{\delta}+3y{\mathcal{G}}-4y{\mathcal{F}}\,,
\
\\
&&
\label{bpn}
b{'}={\sigma}b+3b{\mathcal{G}}\,,
\
\\
&&
\label{lambdapn}
\lambda{'}={\frac{3}{2}}{\lambda}{\mathcal{G}}-{\lambda}{\mathcal{F}}-{\lambda}^2(1-{\Gamma})\,,
\
\\
&&
\label{deltapn}
\delta{'}={\frac{3}{2}}{\delta}{\mathcal{G}}-{\delta}^2(1-{\tau})-{\delta}{\mathcal{F}}\,,
\
\\
&&
\label{zpn}
z{'}=-3z(1+x+y+b(-1+z)-xz-yz)\,,
\end{eqnarray}
where
\fa
&&
\label{F-definition}
\mathcal{F}={\frac{3(x+2y)}{x+6y}}+{\frac{x{\lambda}+3{\delta}y+{\sigma}b}{2(x+6y)}}\,,
\
\\
&&
\label{G-definition}
\mathcal{G}={\frac{(-1+2z)(1+x+y+b(-1+z)-xz-yz)}{-1+z}}\,,
\ffa
and
\fa
\label{parameters}
\Gamma={\frac{\alpha({\frac{d^2{\alpha}}{d{\phi}^2}})}{({\frac{d{\alpha}}{d{\phi}}})^2}}\,,~~~~~
\tau={\frac{\beta({\frac{d^2{\beta}}{d{\phi}^2}})}{({\frac{d{\beta}}{d{\phi}}})^2}}\,,~~~~~
\sigma={\frac{d(ln~V)}{dN}}\,.
\ffa
Notice that here we have introduced the number of e-folding $N\equiv\ln(a/a_0)$ with $a_0$ being the current value of the scale factor and the prime represents the derivative with respect to $N$ in the above equations.

Comparing with the standard FRW cosmology, an additional dimensionless variable $z\equiv\rho_t/\rho_c$ is introduced to describe the system in LQC. The nonzero $z$ represents the LQG effect. Next, we shall treat $\lambda$ and $\delta$ as constant coupling parameters and dynamically changing variables respectively to study the dynamics of the system.

\section{Dynamics with constant coupling parameters}\label{sec-con}

In this section, we consider $\lambda$ and $\delta$ as constant coupling parameters and so we set $\lambda=\lambda_0$ and $\delta=\delta_0$. The the dynamic system reduces to a $4$-dimensional one. A non-trivial $\lambda$ and $\delta$ give the following relation
\begin{eqnarray}
\lambda_0(1-{\Gamma})=\delta_0(1-{\tau})\,.
\end{eqnarray}
From Eqs. \eqref{lambdapn} and \eqref{deltapn}, it is easy to derive the following constraints
\fa
\lambda_0=\frac{\frac{3}{2}\mathcal{G}-\mathcal{F}}{1-\tau}\,,~~~~~
\delta_0=\frac{\frac{3}{2}\mathcal{G}-\mathcal{F}}{1-\Gamma}\,.
\ffa

\subsection{Pure k-essence}

As has pointed out in the introduction, different from the quintessence, the non-canonical kinetic term of k-essence plays a key role in driving the cosmic acceleration. So in this subsection, we first consider the pure k-essence model, i.e., $V(\phi)=0$. In this case, $b=0$ and the system reduces to a $3$-dimensional one.

Also, we are interesting in the k-essence fractional density parameter $\Omega_\phi$ and the effective k-essence equation of state (EoS) parameter $\gamma_\phi$, which reads
\begin{eqnarray}
&&
\label{Omega-v1}
\Omega_\phi=\frac{\rho_\phi}{3H^2}=x+3y\,,
\
\\
&&
\label{EoS-gamma-v1}
\gamma_\phi=1+\omega_\phi=1+\frac{p_\phi}{\rho_\phi}=\frac{2x+4y}{3+3y}\,,
\end{eqnarray}
for $V(\phi)=0$. We see that $\Omega_\phi$ and $\gamma_\phi$ have the same expressions as that in standard FRW cosmology \cite{Chakraborty:2019swx}.
The observations constrain the current values for $\Omega_\phi$ and $\gamma_\phi$ as \cite{Bahamonde:2017ize}
\begin{eqnarray}
&&
\label{Omega-value-current}
\Omega_\phi\approx0.68\,,
\
\\
&&
\label{gamma-value-current}
\gamma_\phi\approx0.05\,.
\end{eqnarray}
Therefor, we have $x_0=-1.309$, $y_0=0.663$ for current universe.

By setting $x'=0$, $y'=0$ and $z'=0$, one obtains the fixed points for this system. We can explore the stability of the fixed points by evaluating the corresponding eigenvalues. Following the strategy outlined in \cite{Copeland:2006wr,Copeland:1997et,Gumjudpai:2005ry}, one can work out the eigenvalues. The corresponding $\Omega_\phi$ and $\gamma_\phi$ are also worked out. These results are summarized in TABLE \ref{tb2-1}.

\begin{table}[htp]
\linespread{1.2}
\centering
\caption{Fixed points for pure k-essence}
\label{tb2-1}
\setlength{\tabcolsep}{1.3 mm}
 \scriptsize
{\begin{tabular}{c|c|c|c|c|c|c|c}
   \hline
    Point              & x                 & y               &z              & Eigenvalues                            & $\Omega_{\phi}$         & $\gamma_{\phi}$          &Stability Condition \\
    \hline\hline
    A.                & 0                  &  $\frac{1}{3}$  &0              &-4,1,$\frac{1}{2}(4-\delta+2\lambda)$    &  1         &$\frac{4}{3}$                             & Saddle point                \\
    \hline
    B.                 & 1  & 0  & 0    &-6,3,$-6+\delta-2\lambda$  & 1   &2   & Saddle point   \\
    \hline
    C.                 & $\frac{1}{2}(\delta-2\lambda-4)$   &$\frac{1}{6}(6-\delta+2\lambda)$        &0           &$-\delta+2\lambda$,$\delta-2\lambda-3$,     &          &     &$2\lambda<\delta<2\lambda+3$ \\
                      &                   &                   &                  &  $\frac{(2\lambda-\delta+4)(2\lambda-\delta+6)}{\delta-2(4+\lambda)}$  &  1     & $\frac{\delta-2\lambda}{3}$             & \\

     \hline
\end{tabular}}
\end{table}

From this table, we can see that the fixed point C is stable if
\fa
\label{cond-v0}
2\lambda<\delta<2\lambda+3\,,
\ffa
is satisfied. Comparing with the case of k-essence in standard FRW cosmology in \cite{Chakraborty:2019swx}, the LQG effect imposes a lower bound on $\delta-2\lambda$. Given the condition \eqref{cond-v0}, one has $\gamma_\phi<0$, i.e., $\omega_\phi>-1$, for which the big rip \cite{Caldwell:2003cr} in later universe is avoided. Notice that to have an accelerated expansion universe at later stage, we have $\gamma_\phi<2/3$, which further leads to a tighter constraint
\fa
\label{cond-v1}
2\lambda<\delta<2\lambda+2\,.
\ffa

\begin{figure}
\center{
\includegraphics[scale=0.58]{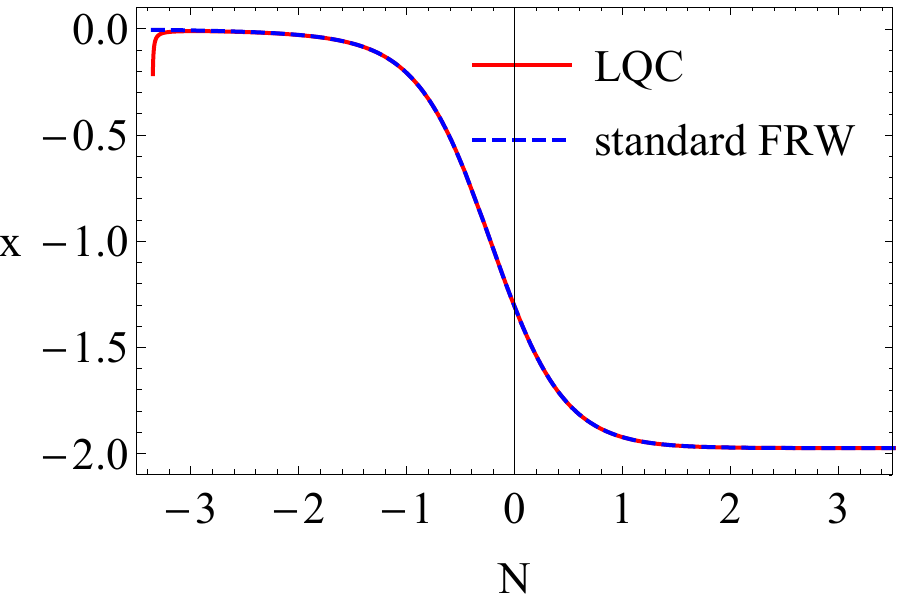}\ \hspace{0.1cm}
\includegraphics[scale=0.55]{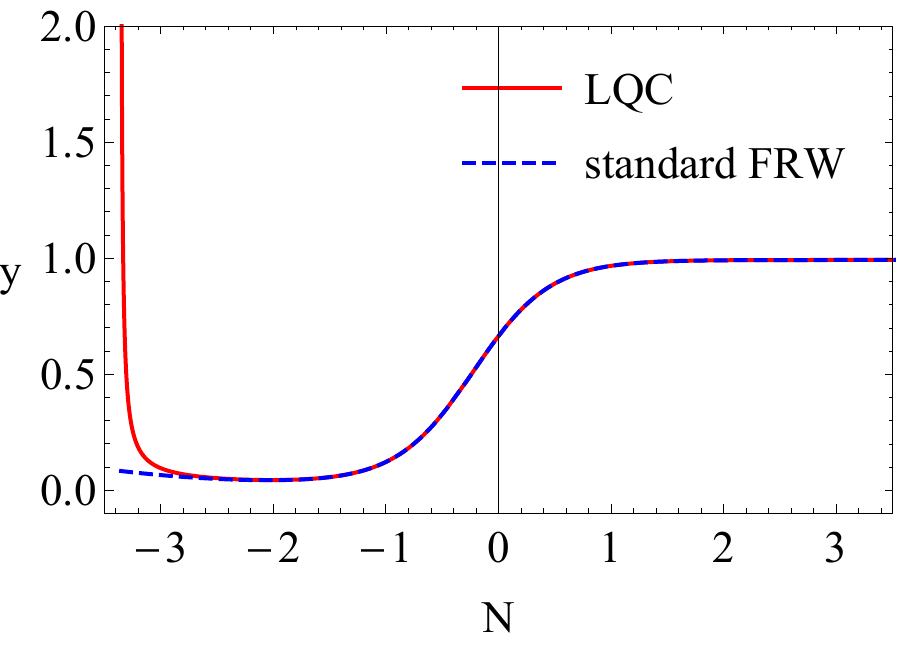}\ \hspace{0.1cm}
\includegraphics[scale=0.4]{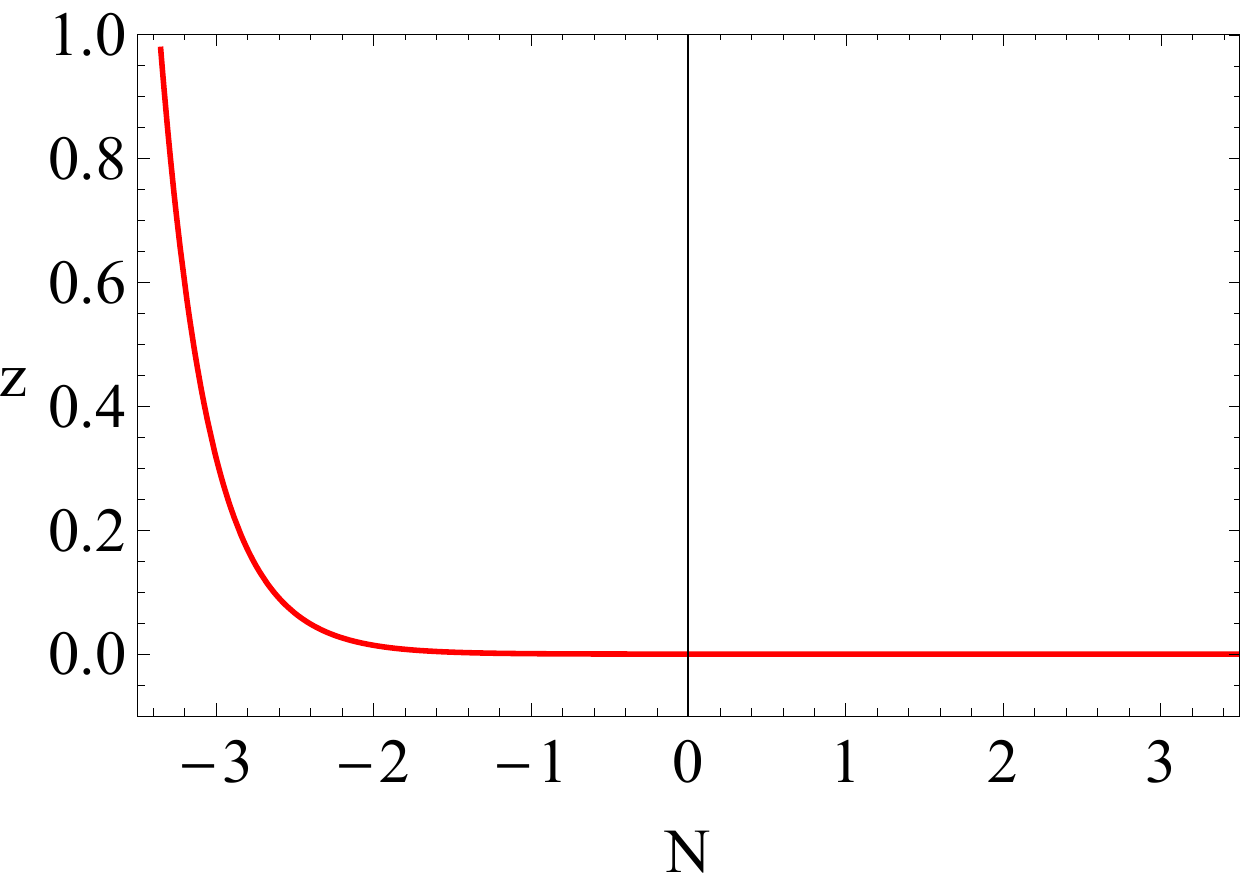}\ \\
\caption{\label{xyzvsN-v0} The evolutions of the system with $N$ for pure k-essence. The red curves are for LQC and the blue dashed curves for the standard FRW cosmology. Here we have chosen $\lambda=2$ and $\delta=4.05$.
}}
\end{figure}
\begin{figure}
\center{
\includegraphics[scale=0.8]{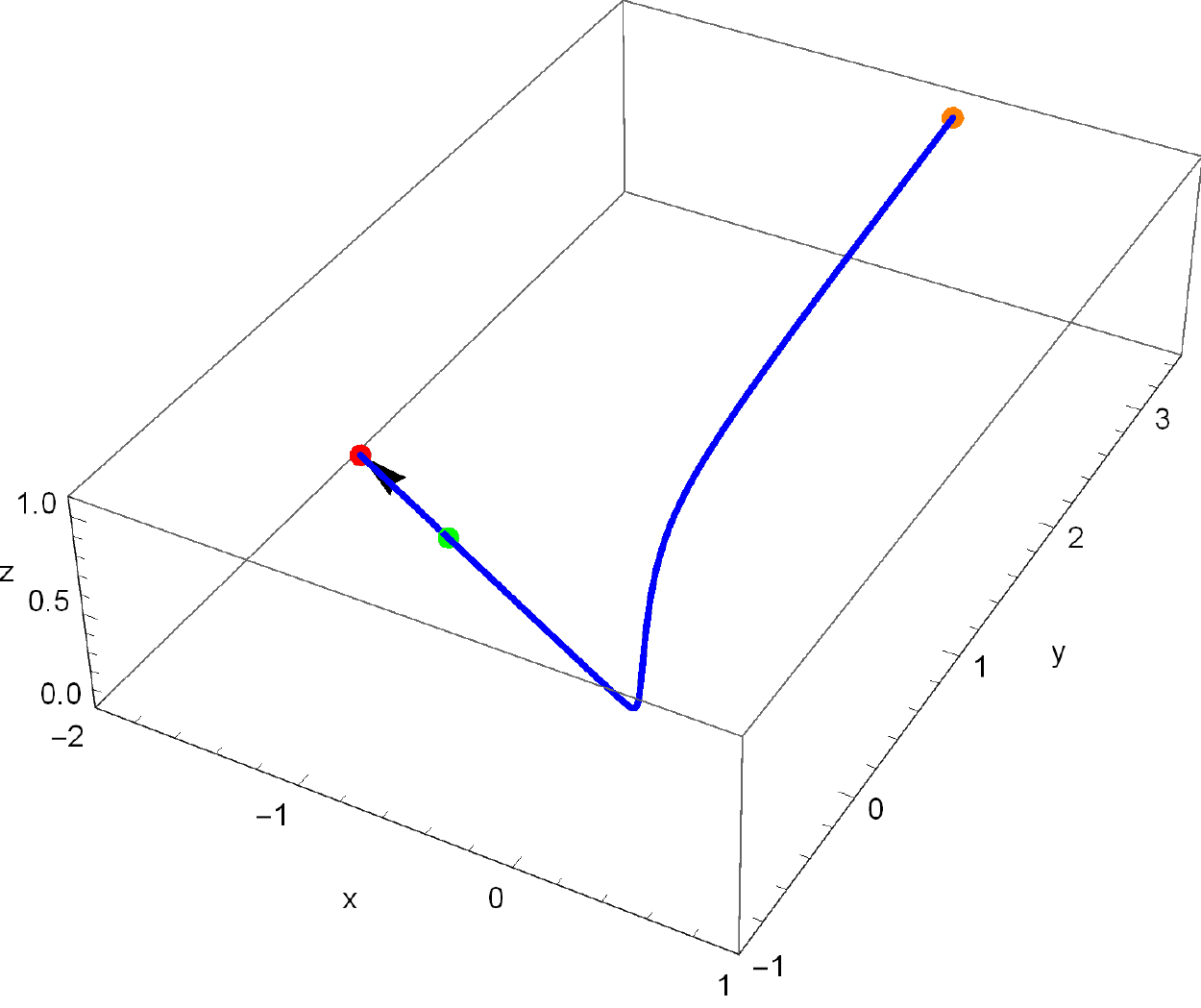}\ \hspace{0.7cm}
\caption{\label{p3dlqcv0} The parameter plot of $x$, $y$ and $z$ for pure k-essence. The blue curve exhibits the time evolution from past (orange point) to the attractor (red point). The green point stands for the present universe. Here we have chosen $\lambda=2$ and $\delta=4.05$.
}}
\end{figure}

Now, we turn to study the evolution of the system with N. To solve the EOMs of this system, we take the initial condition to satisfy the current observation constrain, i.e., Eqs. \eqref{Omega-value-current} and \eqref{gamma-value-current}. Besides, we assume that $z$ is small at the current universe. The evolutions of $x$, $y$ and $z$ as the function of $N$ are shown in FIG.\ref{xyzvsN-v0}. The red curves are for LQC and the blue dashed curves for the standard FRW cosmology. Notice that the parameters $\lambda$ and $\delta$ are chosen as $\lambda=2$ and $\delta=4.05$, which satisfy the stability condition \eqref{cond-v1}.

From this figure, we find that for most of the time of the universe evolution, $x$ and $y$ in LQC are the same as that in the standard FRW cosmology. Only in the early epoch of the universe, $x$ rapidly decreases and $y$ increases as time turns back. On the other hand, $z$ almost vanishes for most of the time of the universe evolution. But as time turns back, it rapidly increases in the early epoch of the universe. These phenomena indicate that LQG effect plays an important role only in the early epoch of the universe. Also, we show the parameter plot of $x$, $y$ and $z$ in FIG.\ref{p3dlqcv0}. Indeed, the universe finally evolutes into the stability attractor solution (point C, red point in FIG.\ref{p3dlqcv0}).

\begin{figure}
\center{
\includegraphics[scale=0.85]{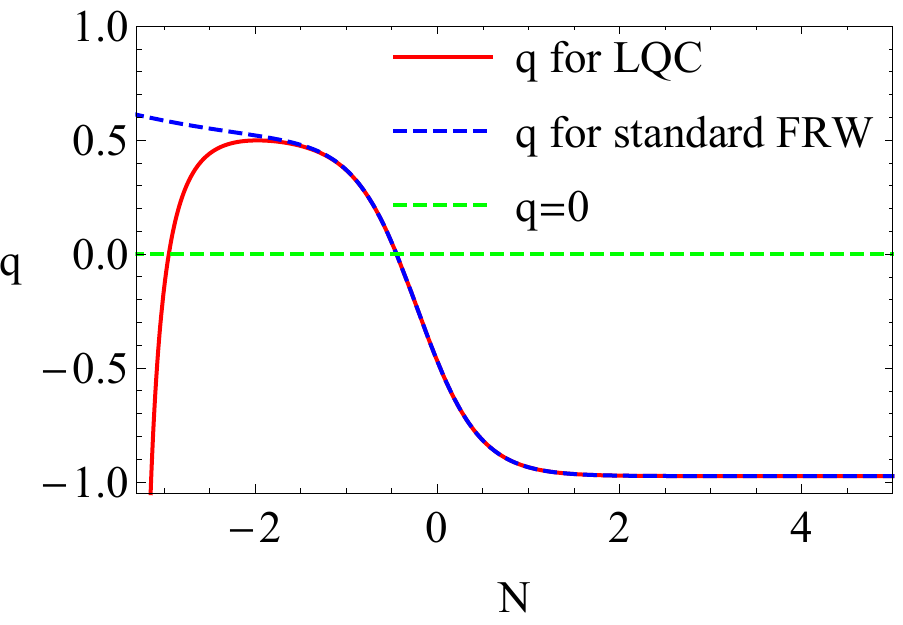}\ \\
\caption{\label{qvsN-v0} The evolutions of the deceleration parameter $q$ with $N$ for pure k-essence. The red curve is for LQC and the blue dashed curve for the standard FRW cosmology. Here we have chosen $\lambda=2$ and $\delta=4.05$.
}}
\end{figure}

Further we plot the evolutions of the deceleration parameter $q$ with $N$ (FIG.\ref{qvsN-v0}). We see that in the early epoch, the universe undergoes a so-called super-inflation stage due to the LQG effect \cite{Fu:2008gh,Trojanowski:2020xza,Pongkitivanichkul:2020txi,Pacif:2020hai}. After that, the universe enters into a decelerated phase, which is almost the same as that for the standard FRW cosmology. And then, the universe changes from this decelerated phase to an accelerated expansion stage where the LQG effect is diluted.

\begin{figure}
\center{
\includegraphics[scale=0.82]{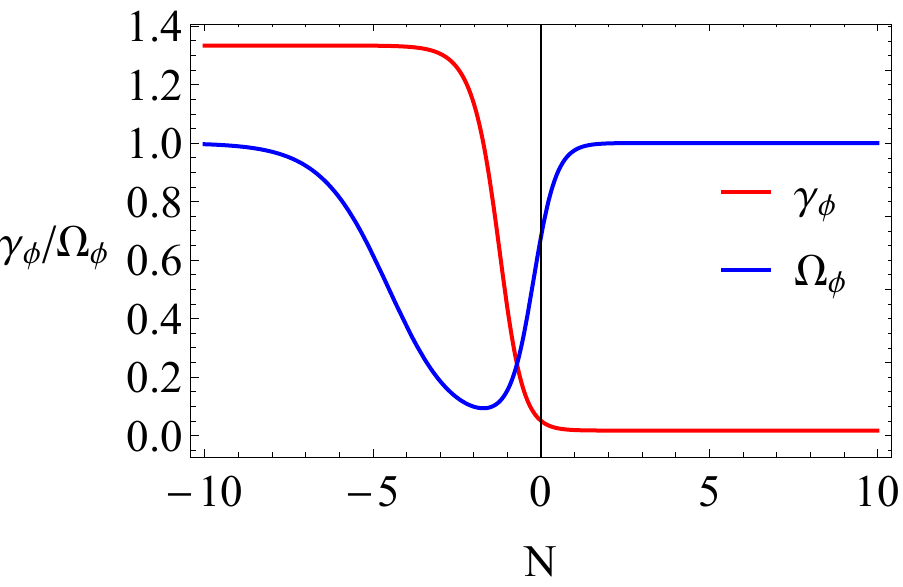}\ \hspace{0.8cm}
\includegraphics[scale=0.8]{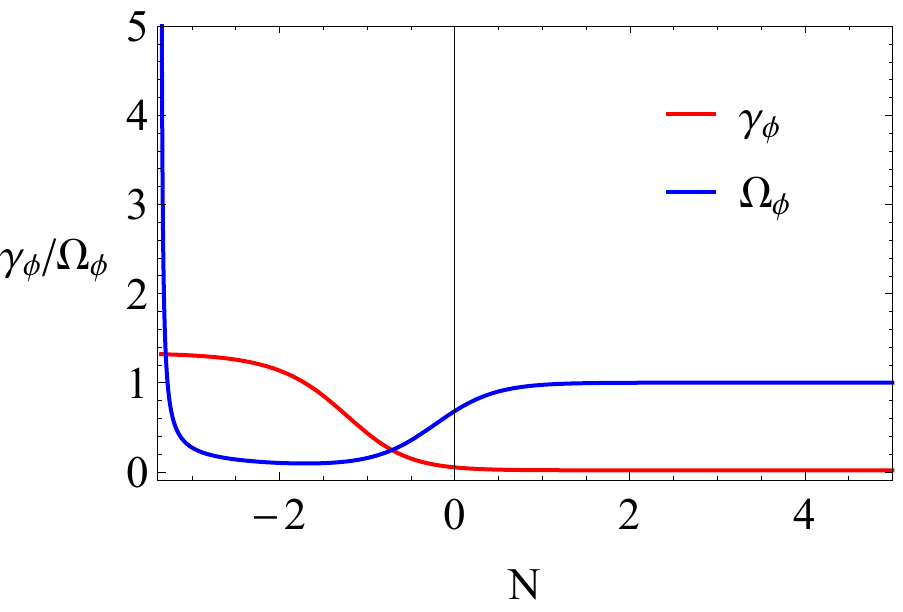}\ \\
\caption{\label{gammavsN-lqc-v0} The evolutions of $\gamma_\phi$ and $\Omega_\phi$ with $N$ for pure k-essence. Left plot is for standard FRW cosmology and right plot for LQC. Here we have chosen $\lambda=2$ and $\delta=4.05$.
}}
\end{figure}

In FIG.\ref{gammavsN-lqc-v0}, we show the evolution of $\gamma_\phi$ and $\Omega_\phi$ with $N$ in the standard FRW cosmology (left) and LQC (right), respectively. Again, LQC effect plays a key role only in the early universe and is diluted as the time evolutes. Finally, the universe evolves into the stable attractor with $\gamma_\phi=0$ and $\Omega_\phi=1$, which means the universe enters into a scalar field dominated one.

Finally, we would like to point out that the fixed points A and B are not stable fixed points but are the saddle ones because there are at least one negative eigenvalue for them (see TABLE \ref{tb2-1}).

\subsection{K-essence with nonzero potential}

\begin{table}[htp]
\linespread{1.2}
\centering
\caption{Fixed points for k-essence with non-zero potential}
\label{tb2-2}
\setlength{\tabcolsep}{0.2 mm}
 \scriptsize
{\begin{tabular}{c|c|c|c|c|c|c|c|c}
   \hline
    Point    & x & y&z&b &Eigenvalues& $\Omega_{\phi}$ & $\gamma_{\phi}$ &Stability condition \\
    \hline\hline
    A        & 0 &$-\frac{\sigma}{12}$& 0 &$\frac{4+\sigma}{4}$ & $\frac{1}{2}(-\delta+2\lambda-\sigma)$,$\sigma$,$-\sigma-3$,$-4-\sigma$ &  1 & ${-\frac{\sigma}{3}}$ &$2\lambda<\delta+\sigma$ and $\sigma>-3$ \\
    \hline
    B      & $-\frac{\sigma}{6}$  & 0  & 0&$\frac{6+\sigma}{6}$     & $\sigma$,$-3-\sigma$,$-6-\sigma$ ,$\delta-2\lambda+\sigma$ & 1   &${-\frac{\sigma}{3}}$& $2\lambda>\delta+\sigma$ and $\sigma>-3$\\
    \hline
    C        & $\frac{1}{2}(\delta-2(2+\lambda))$  & $\frac{1}{6}(6-\delta+2\lambda)$  & 0   & 0&$-\delta+2\lambda$,$\frac{\lambda^2-2\delta(5+2\lambda)+4(6+5\lambda+\lambda^2)}{\delta-2(4+\lambda)}$,$\delta-2\lambda-3$,$\delta-2\lambda+\sigma$ & 1 &${\frac{\delta-2\lambda}{3}}$
    & $2\lambda>\delta+\sigma$ and $\sigma>-3$   \\
    \hline
    D         & 1  & 0  & 0  &  0& -4,1,$\frac{1}{2}(4-\delta+2\lambda)$,$4+\sigma$  & 1  &${\frac{4}{3}}$     &unstable \\
    \hline
    E        & 0 &  $\frac{1}{3}$  &0& 0&{-6,3,$\delta-6-2\lambda$},$6+\sigma$&1 &1&unstable\\
     \hline
\end{tabular}}
\end{table}
In this subsection, we study the dynamics of k-essence with nonzero potential. $V(\phi)\neq0$ leads to $b\neq 0$ and so the dimension of this dynamical system becomes a $4$-dimensional one. Then the scalar field fractional energy density $\Omega_\phi$ and the EOS parameter $\gamma_\phi$ become
\begin{eqnarray}
&&
\label{Omega-v1}
\Omega_\phi=x+3y+b\,,
\
\\
&&
\label{EoS-gamma-v1}
\gamma_\phi=\frac{2x+4y}{3+3y+b}\,.
\end{eqnarray}
The values of $\Omega_\phi$ and $\gamma_\phi$ of the current universe (Eqs. \eqref{Omega-value-current} and \eqref{gamma-value-current}) gives the initial condition space as
\begin{eqnarray}
&&
\label{inicon1}
x+3y+b=0.68\,,
\
\\
&&
\label{inicon2}
2x+4y=0.034\,.
\end{eqnarray}
Given the initial condition, we can determine the evolutions of $x$, $y$, $z$ and $b$ with $N$. Notice that the initial condition space is the same as that in the standard FRW cosmology. For the detailed discussions, please refer to \cite{Chakraborty:2019swx}.

Following the same procedure above, we work out the fixed points and the stability conditions, which are shown in TABLE \ref{tb2-2}. From this table, we can see that the points D and E are the unstable points. We do not discuss these unstable points.

\begin{figure}
\center{
\includegraphics[scale=0.68]{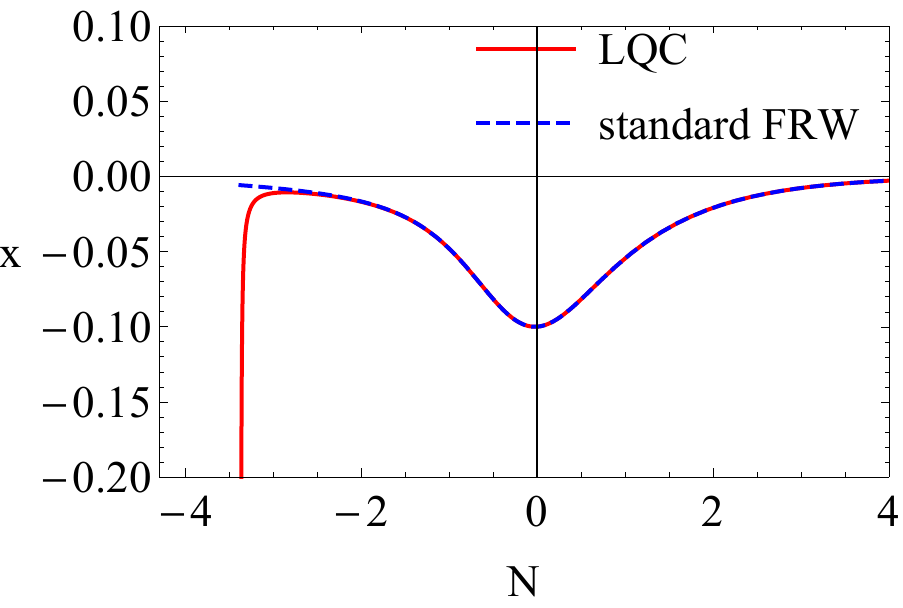}\ \hspace{0.5cm}
\includegraphics[scale=0.68]{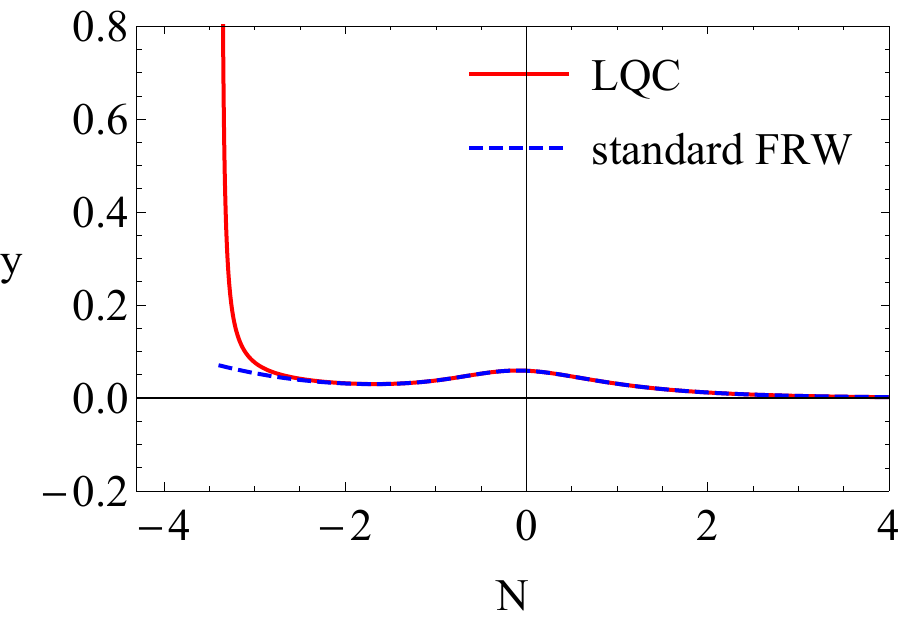}\ \\
\includegraphics[scale=0.48]{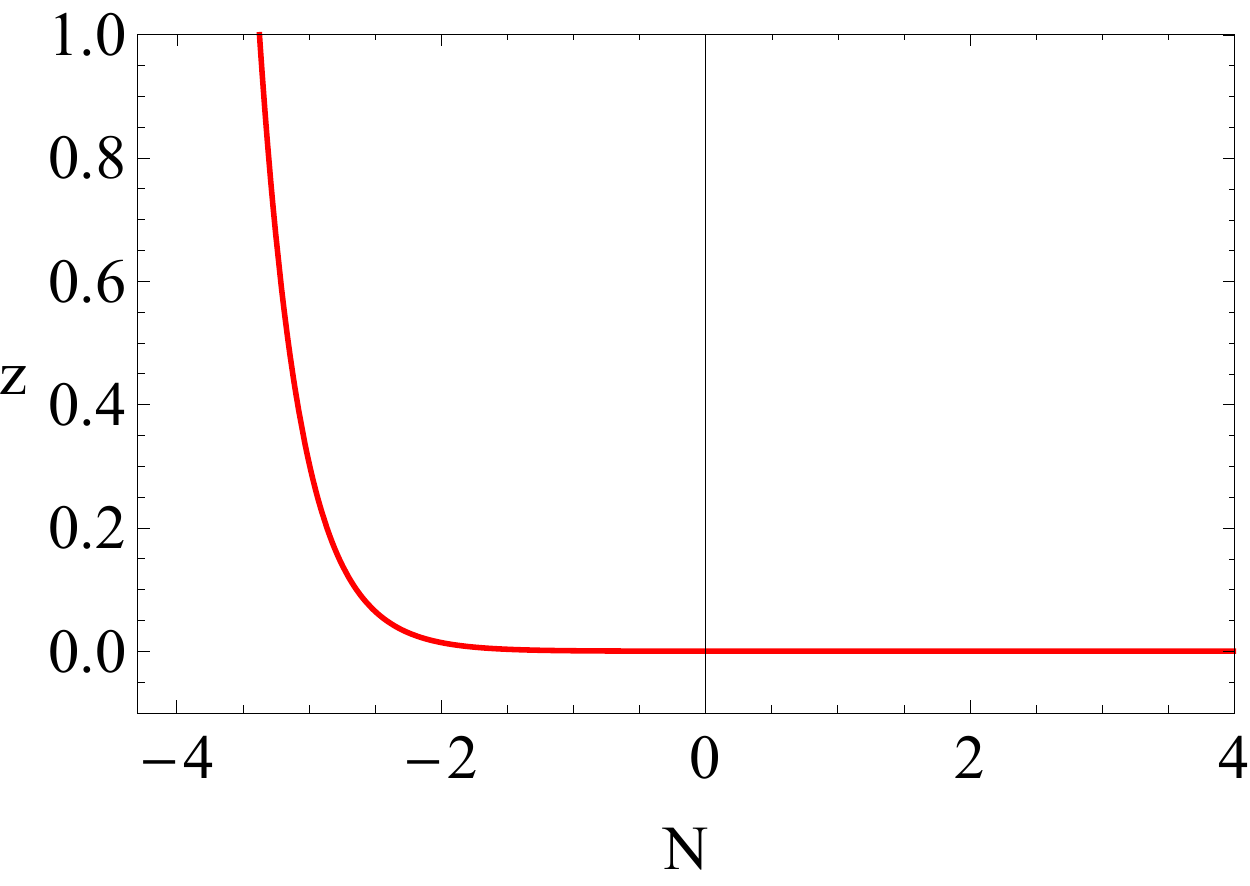}\  \hspace{0.5cm}
\includegraphics[scale=0.68]{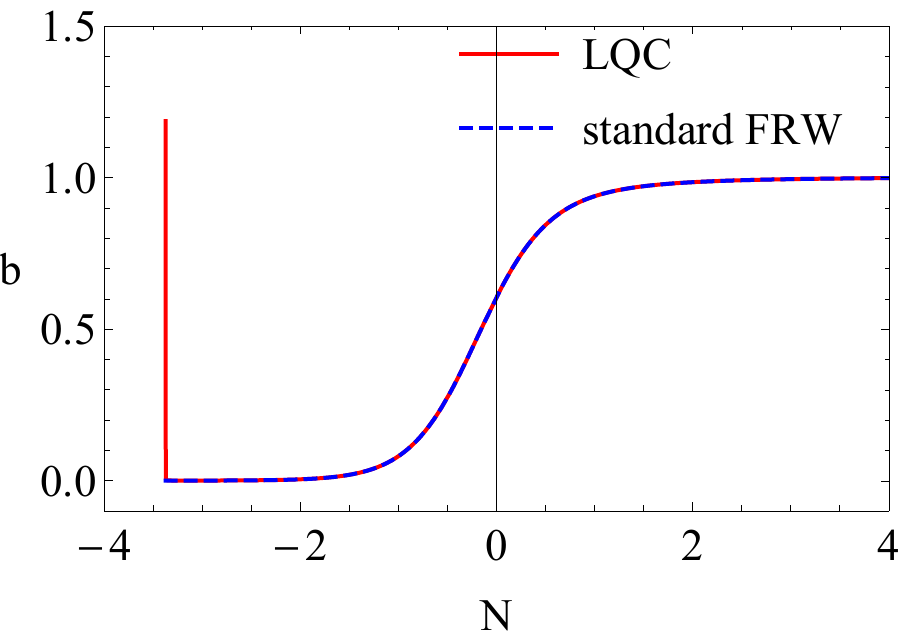}\ \\
\caption{\label{xyzbvsN-v1} The evolutions flowing to fixed point A for k-essence with nonzero potential. The red curves are for LQC and the blue dashed curves for the standard FRW cosmology. Here we have chosen $\lambda=1$ and $\delta=3$ and $\sigma=0$.
}}
\end{figure}
If the condition $2\lambda<\lambda+\sigma$ with $\sigma>-3$ is satisfied, A is a stable fixed point. We show the evolutions flowing to the fixed point A in FIG.~\ref{xyzbvsN-v1}. Here we have fixed $\sigma=0$, which corresponds to the case of $\omega_\phi=-1$ \footnote{For $\sigma<0$, we have similar results. Notice that we do not consider the case of $\sigma>0$, which cross the phantom field divide.}. Regardless of the LQC or the standard FRW universe, the systems flow to the same fixed point $A$. We note that as the time evolves, the linear and quadratic kinetic energy terms $x$ and $y$ reduce to zero, but the potential term $b$ increases and tends to the maximum value of $b=1$. It suggests that as the quintessence dark energy model, the potential plays the dominant role in driving the cosmic acceleration in the later epoch of the universe. It is different from that for the pure k-essence, for which the kinetic energy terms play the role of driving the cosmic acceleration. As the case of pure k-essence studied above, only in the early stage of the universe, the LQG effects play an important role. In most of the evolution time of the universe, the evolution of the system in LQC are basically consistent with that in the standard FRW cosmology \cite{Ito:2012,Lin:2020fue,LQGRovelli,Guo:2003eu}.

\begin{figure}
\center{
\includegraphics[scale=0.85]{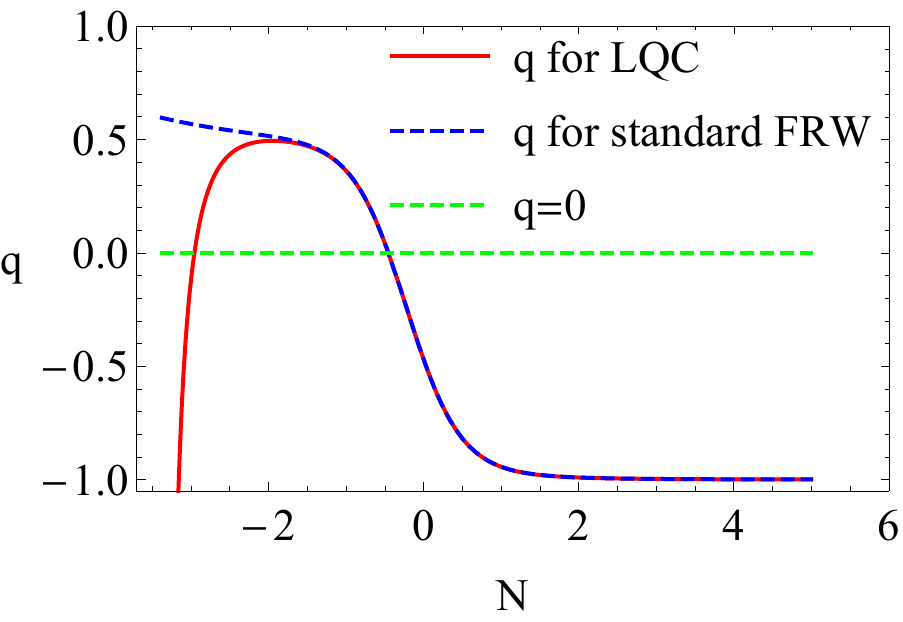}\ \\
\caption{\label{qvsN-v1} The evolutions of the deceleration parameter $q$ with $N$ for k-essence with nonzero potential. The red curve is for LQC and the blue dashed curve for the standard FRW cosmology. Here we have chosen $\lambda=1$ and $\delta=3$ and $\sigma=0$.
}}
\end{figure}
\begin{figure}
\center{
\includegraphics[scale=0.82]{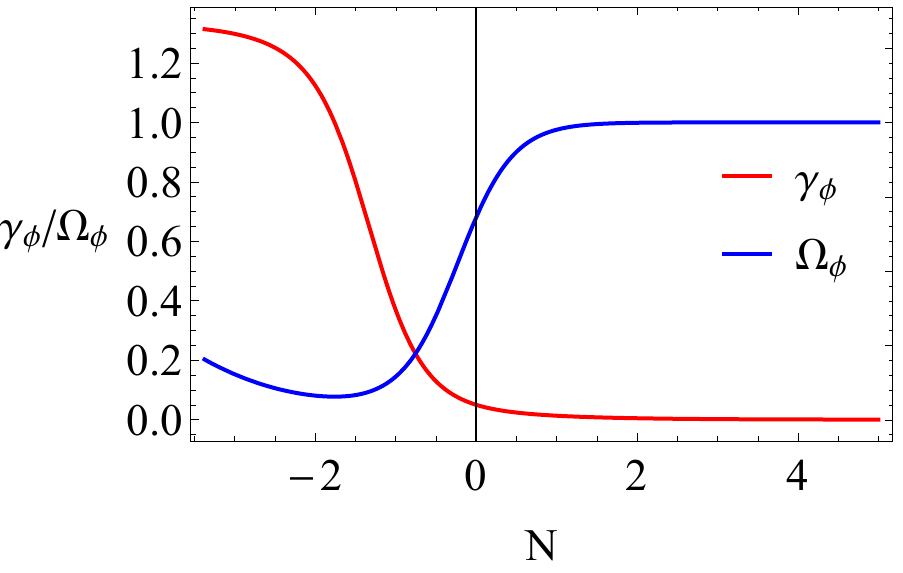}\ \hspace{0.8cm}
\includegraphics[scale=0.8]{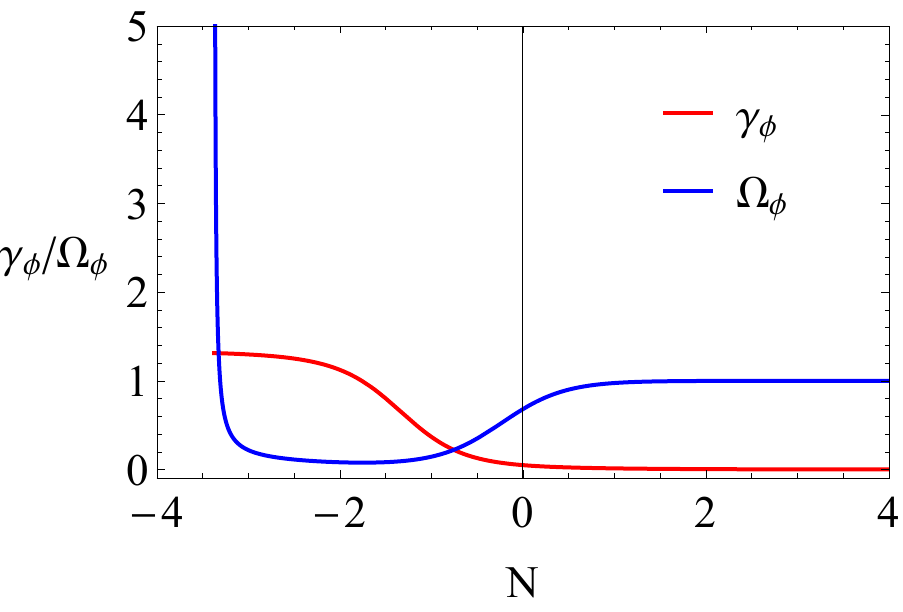}\ \\
\caption{\label{gammavsN-lqc-v1} The evolutions of $\gamma_\phi$ and $\Omega_\phi$ with $N$ for k-essence with nonzero potential. Left plot is for standard FRW cosmology and right plot for LQC. Here we have chosen $\lambda=1$ and $\delta=3$ and $\sigma=0$.
}}
\end{figure}

The corresponding $q$, and $\gamma_\phi$/$\Omega_\phi$ are also shown in FIG.~\ref{qvsN-v1} and FIG.~\ref{gammavsN-lqc-v1}, respectively. From the two figures, we can see that the LQG effect plays an import role in the early epoch of the universe such that the universe undergoes a super-inflation stage. And then, the LQG effect is diluted and the evolution of the universe is almost the same as that in the standard FRW universe. Finally, the universe flows to the scalar field dominated one.
\begin{figure}
\center{
\includegraphics[scale=0.58]{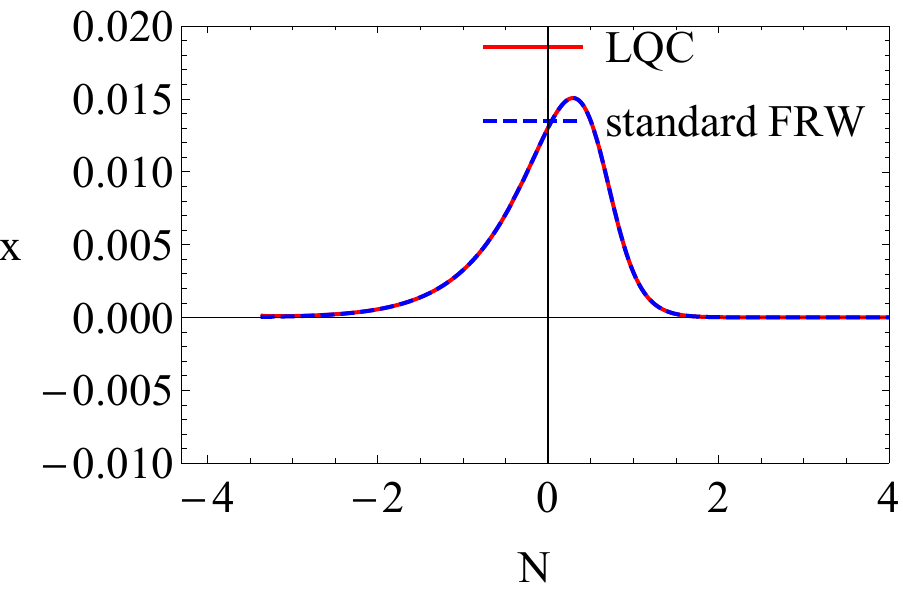}\ \hspace{0.1cm}
\includegraphics[scale=0.58]{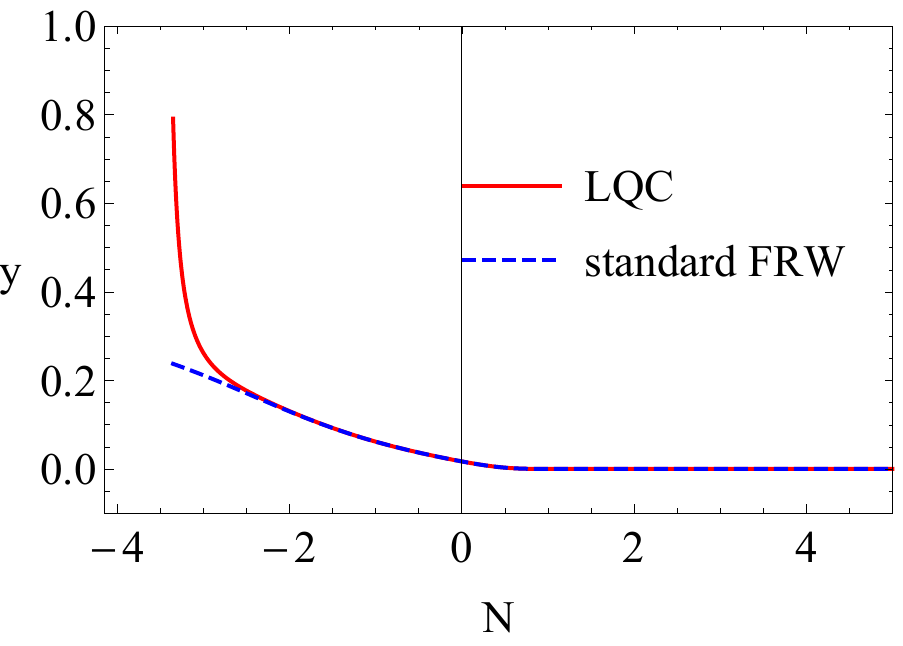}\ \\
\includegraphics[scale=0.4]{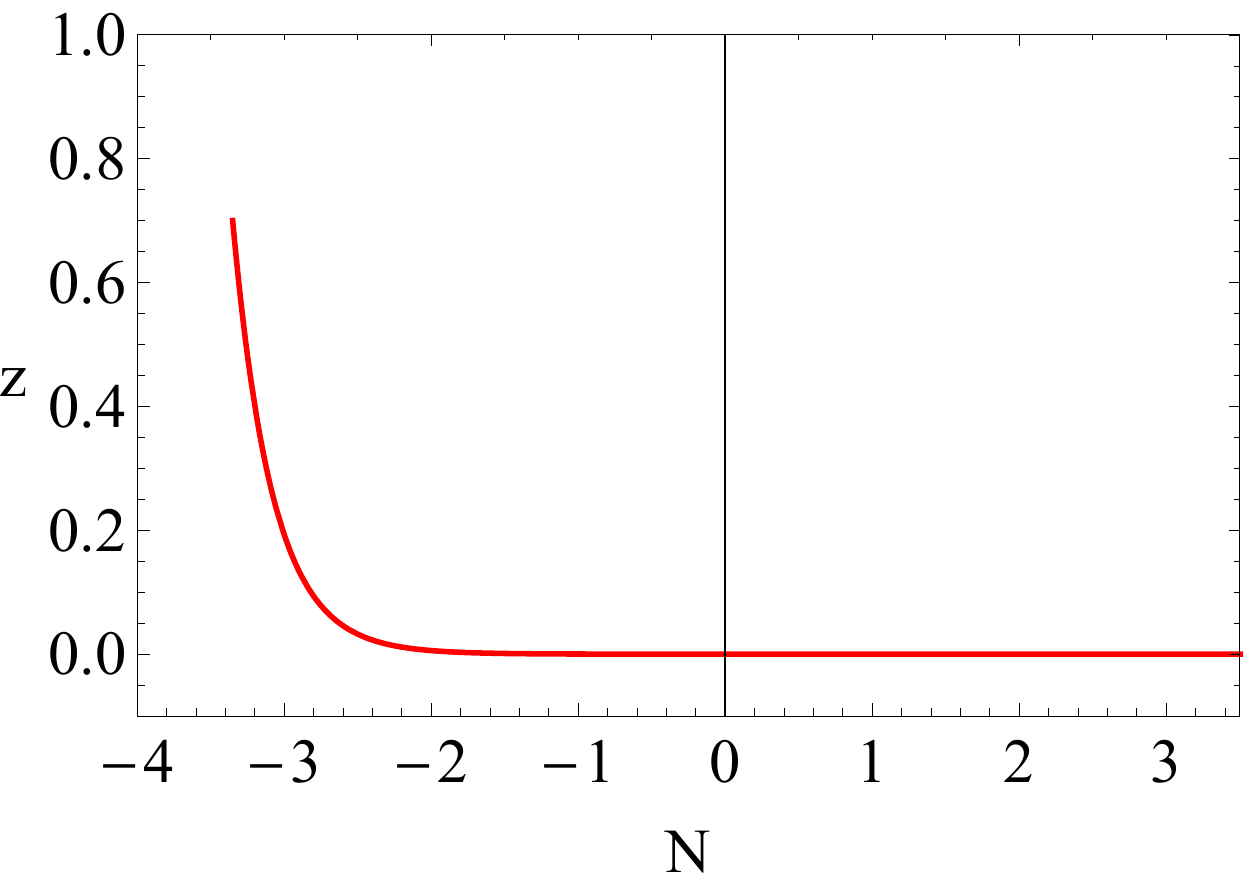}\  \hspace{0.1cm}
\includegraphics[scale=0.58]{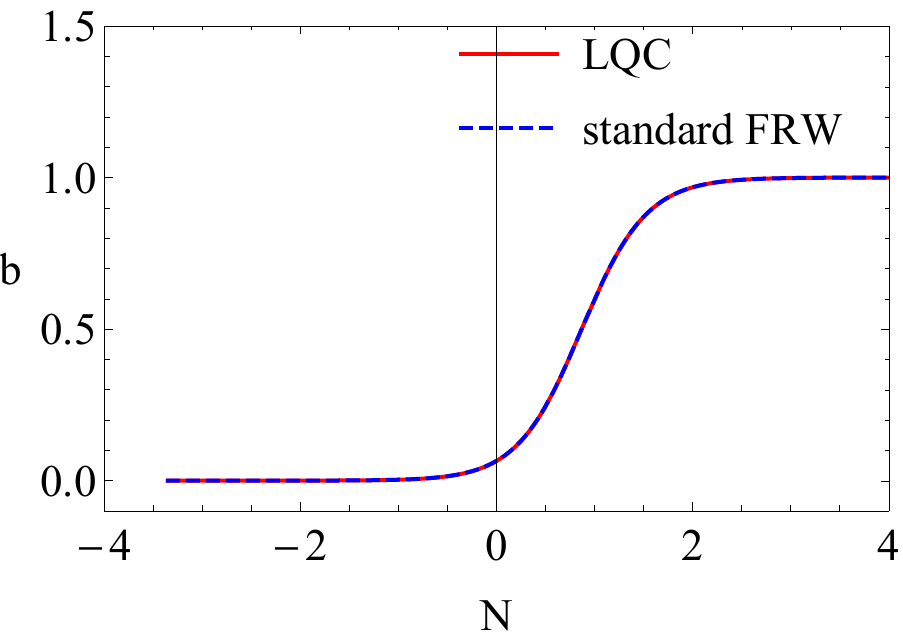}\ \\
\caption{\label{xyzbvsN-v1b} The evolutions of $x$, $y$, $z$ and $b$ with $N$ for k-essence with nonzero potential. The red curves are for LQC and the blue dashed curves for the standard FRW cosmology. Here we have chosen $\lambda=1$ and $\delta=1$ and $\sigma=0$. The initial condition is that $x_0>0$.
}}
\end{figure}
\begin{figure}
\center{
\includegraphics[scale=0.58]{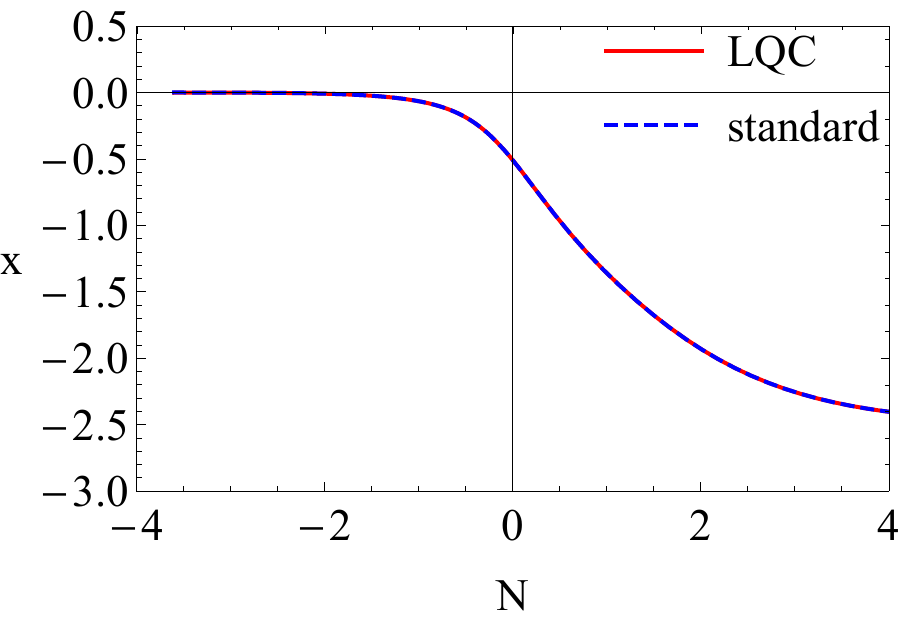}\ \hspace{0.1cm}
\includegraphics[scale=0.58]{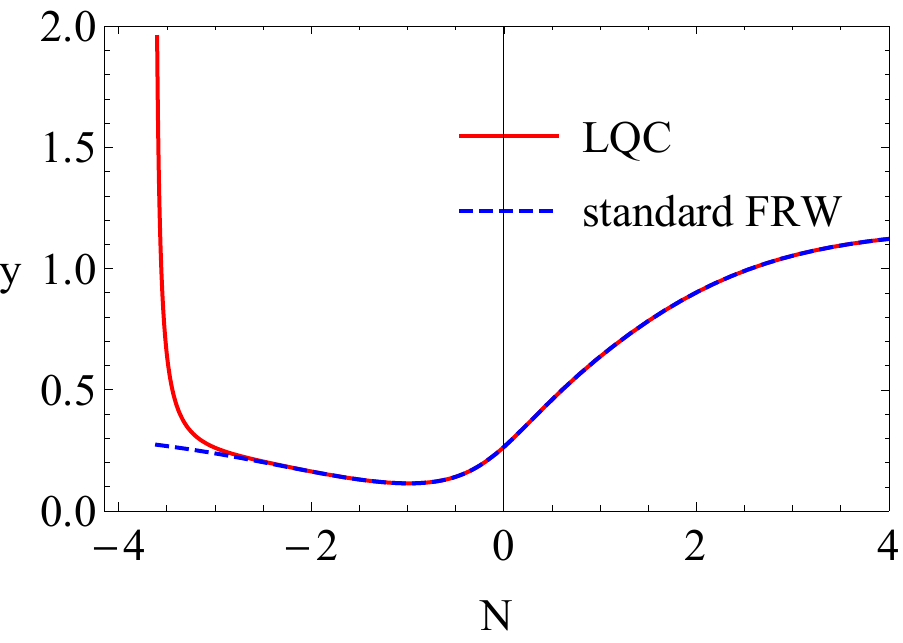}\ \\
\includegraphics[scale=0.4]{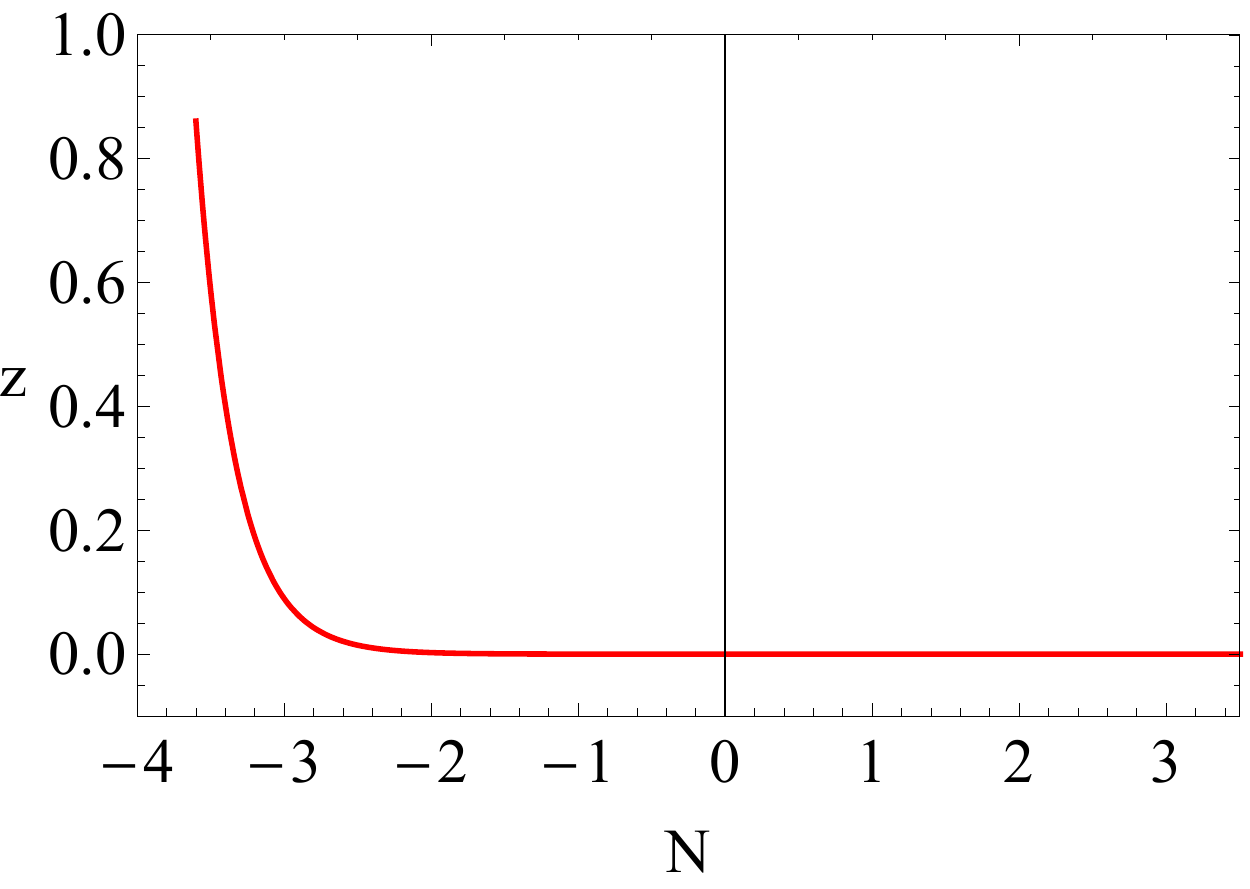}\  \hspace{0.1cm}
\includegraphics[scale=0.58]{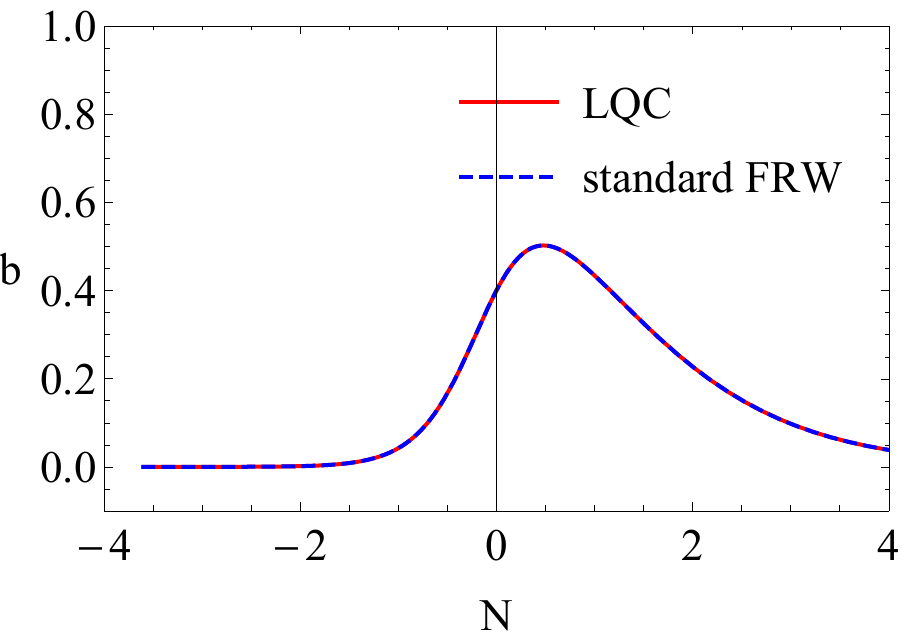}\ \\
\caption{\label{xyzbvsN-v1a} The evolutions of $x$, $y$, $z$ and $b$ with $N$ for k-essence with nonzero potential. The red curves are for LQC and the blue dashed curves for the standard FRW cosmology. Here we have chosen $\lambda=1$ and $\delta=1$ and $\sigma=0$. The initial condition is that $x_0<0$.
}}
\end{figure}
\begin{figure}
\center{
\includegraphics[scale=0.75]{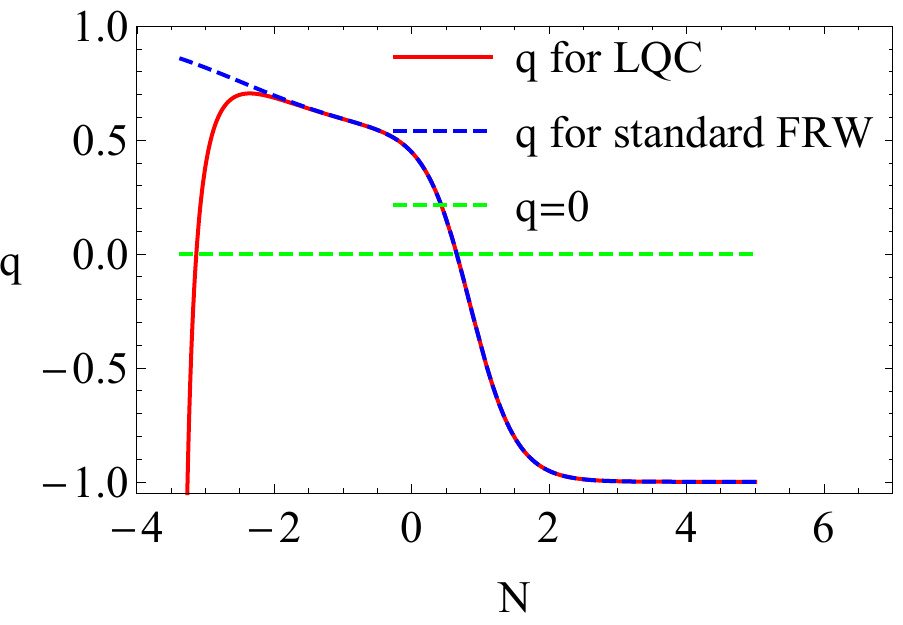}\ \hspace{0.5cm}
\includegraphics[scale=0.75]{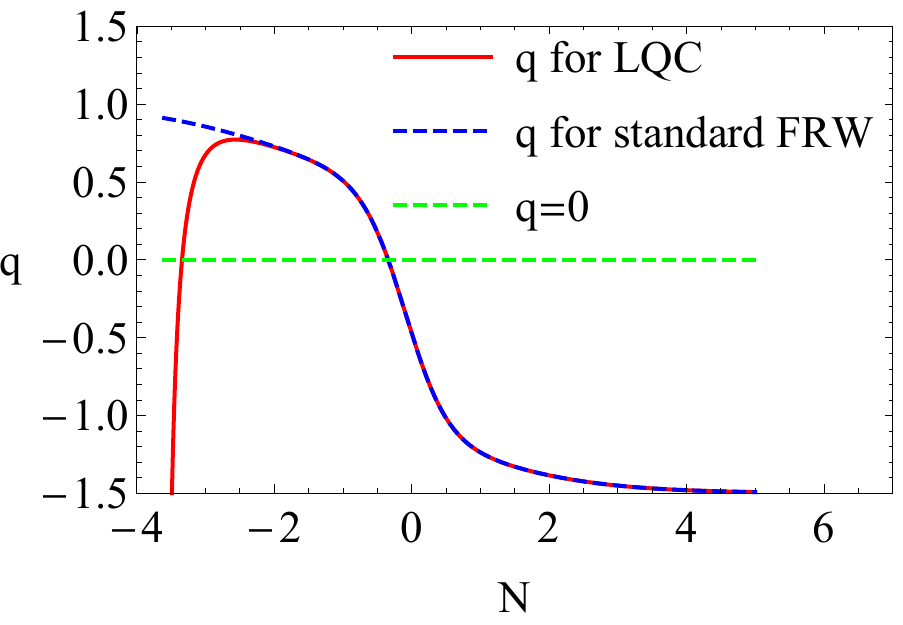}\ \\
\caption{\label{qvsN-v11a} The evolutions of the deceleration parameter $q$ with $N$ for k-essence with nonzero potential. The red curve is for LQC and the blue dashed curve for the standard FRW cosmology. Here we have chosen $\lambda=1$ and $\delta=1$ and $\sigma=0$. Left plot for $x_0>0$ and right for $x_0<0$.
}}
\end{figure}

When $2\lambda>\delta+\sigma$ with $\sigma>-3$, the universe can flow to different fixed points $B$ and $C$ depending on the initial conditions. When the initial condition $x_0$ belongs to the region $x_0>0$, the universe flows to the fixed point B, which is a potential energy dominant case (see FIG.~\ref{xyzbvsN-v1b}). If $x_0<0$, then the universe shall evolve into the fixed point C that a kinetic energy dominant case (see FIG.~\ref{xyzbvsN-v1a}). And then, we also plot the evolutions of the deceleration parameter $q$ with $N$ for the initial conditions $x_0>0$ and $x_0<0$ in FIG.~\ref{qvsN-v11a}, respectively. We find that for the different initial conditions, the evolutions of the deceleration parameter $q$ are qualitatively consistent. That is to say, due to the LQG effect, the universe firstly undergoes a super-inflation stage. Soon afterwards, the universe enters into a decelerated phase and finally under the driving of scalar field, it enters into the accelerated expansion stage.

\section{Dynamically changing $\lambda$ and $\delta$}\label{sec-dynmics}

In the previous section, $\lambda$ and $\delta$ are constant. Next, we study the evolution of the system with dynamically changing $\lambda$ and $\delta$. We shall mainly study the case of pure k-essence. For the k-essence with nonzero potential, we only present a brief discussion.

The system of the pure k-essence with dynamically changing $\lambda$ and $\delta$ is a $5$-dimensional system. Following the same procedure above, we can work out the fixed points and the corresponding eigenvalues, the parameters $\Omega_\phi$ and $\gamma_\phi$ in TABLE \ref{tb2-3}. The stability conditions are also summarized in this table.

\begin{table}[htp]
\linespread{1.2}
\centering
\caption{Fixed points for pure k-essence with dynamically changing $\lambda$ and $\delta$}
\label{tb2-3}
\setlength{\tabcolsep}{1.0 mm}
 \scriptsize
{\begin{tabular}{c|c|c|c|c|c|c|c|c|c}
   \hline
    Point    & x & y&z &$\lambda$&$\delta$& $\Omega_{\phi}$ & $\gamma_{\phi}$ &Eigenvalues&Stability Condition \\
    \hline\hline
    A        & 0 &$\frac{1}{3}$   &0  &0                                              &0                      &1 &$\frac{4}{3}$ &${-4,2,1,1,1}$                                                                   &Unstable\\
    \hline
    B        & 0 &$\frac{1}{3}$   &0  &0                                              &$\frac{4}{5-4\tau}$    &1 &$\frac{4}{3}$ &$\frac{4(-1+\tau)}{-5+4\tau}$,$\frac{8(-1+\tau)}{-5+4\tau}$,-4,-1,1              &Unstable\\
    \hline
    C        & 0 &$\frac{1}{3}$   &0  &$\frac{1}{1-\Gamma}$                           &0                      &1 &$\frac{4}{3}$ &-4,-1,1,1,$\frac{-3+2\Gamma}{-1+\Gamma}$                                         &Unstable \\
    \hline
    D        & 0 &$\frac{1}{3}$   &0  &$\frac{4(-1+\tau)}{(-1+\Gamma)(-5+4\tau)}$    &$\frac{4}{5-4\tau}$    &1 &$\frac{4}{3}$ &$-\frac{4(-1+\tau)}{-5+4\tau}$,$\frac{4(-3+2\Gamma)(-1+\tau)}{(-1+\Gamma)(-5+4\tau)}$,-4,-1,1
    &Unstable   \\
    \hline
    E        & -2 &1               &0  &0                                              &0                      &1 &0             &{-3,-3,0,0,0}                                                                     &Saddle point \\
     \hline
    F        &1   &0              &0   &0                                              &0                      &1 &2             &{-6,-6,3,0,0}                                                                    &Unstable\\
    \hline
\end{tabular}}
\end{table}
\begin{figure}
\center{
\includegraphics[scale=0.5]{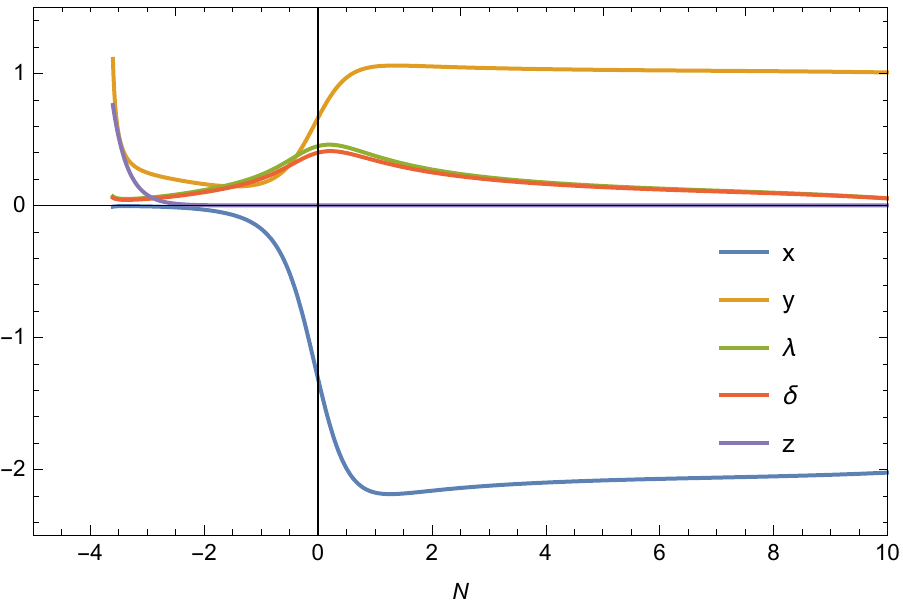}\ \hspace{0.2cm}
\includegraphics[scale=0.5]{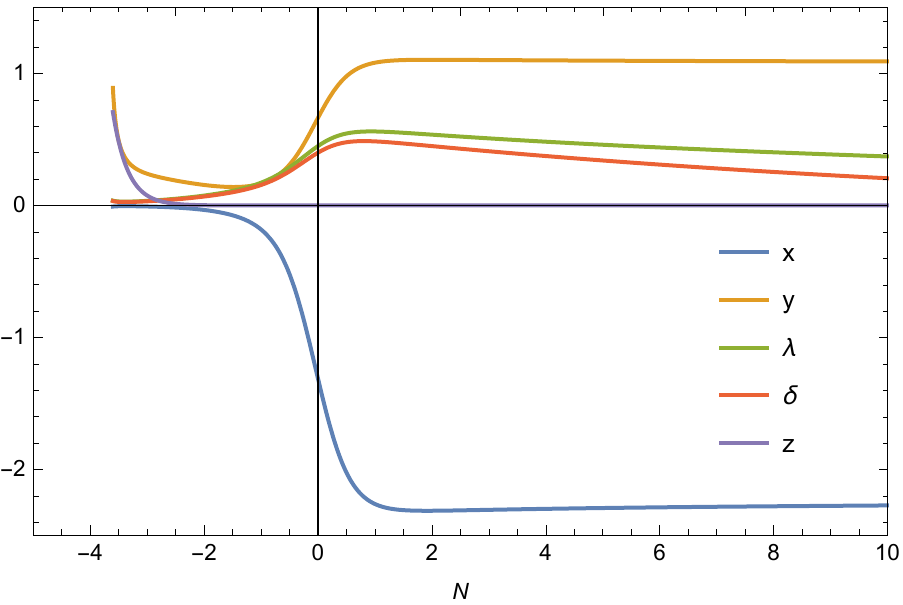}\ \hspace{0.2cm}
\includegraphics[scale=0.5]{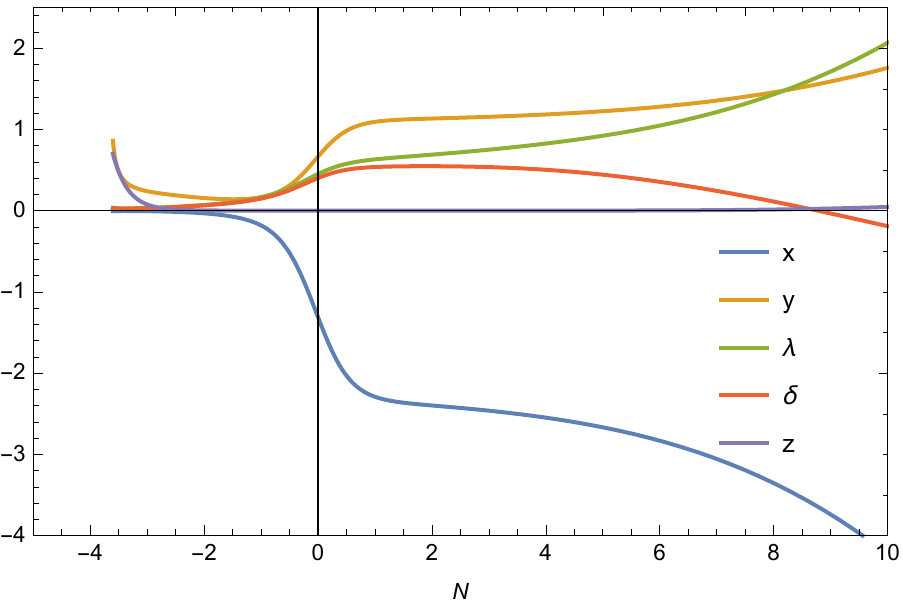}\ \\
\caption{\label{xqvsN-v7c} The evolutions of the system with N for pure k-essence with dynamical changing $\lambda$ and $\delta$.
From left to right, $\Gamma=\tau=0.5,1.4,1.6$.
}}
\end{figure}
\begin{figure}
\center{
\includegraphics[scale=0.56]{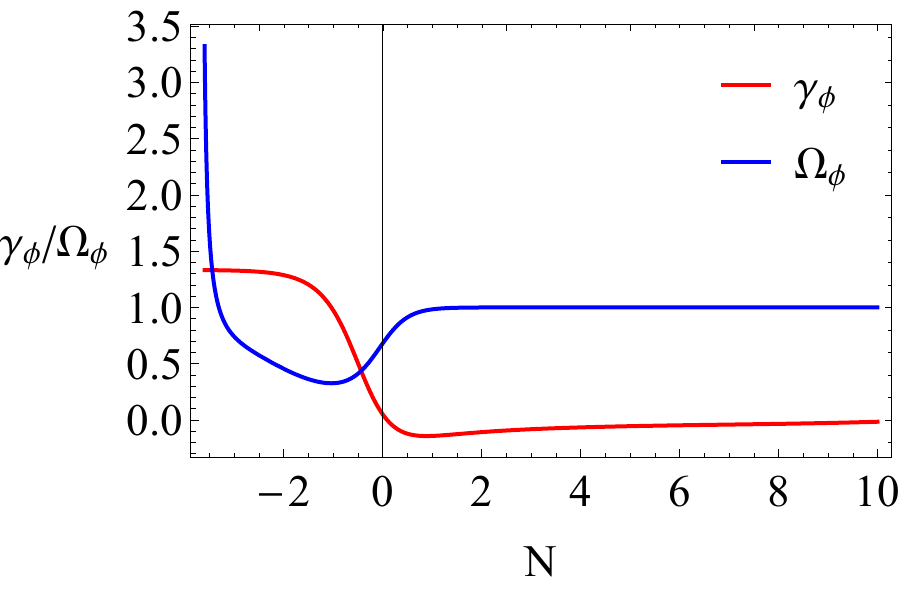}\ \hspace{0.2cm}
\includegraphics[scale=0.56]{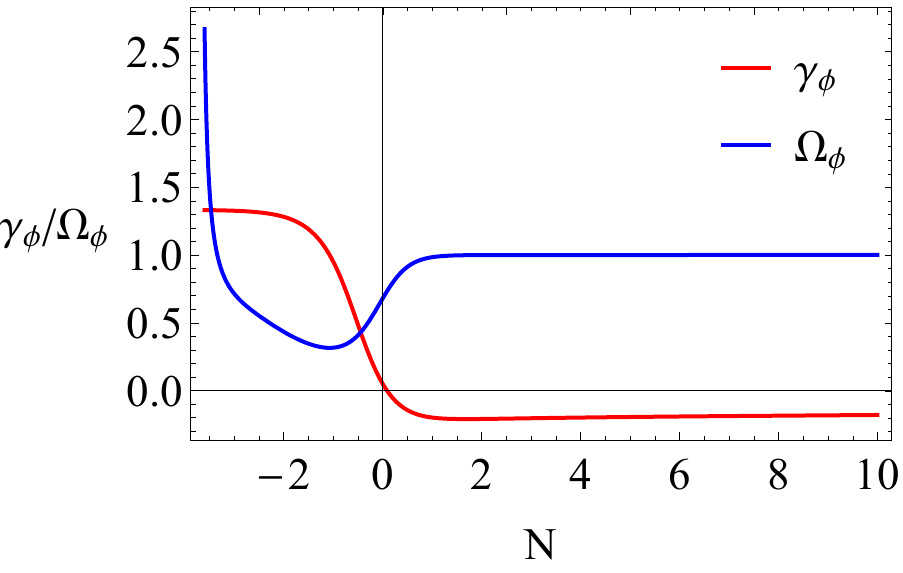}\ \hspace{0.2cm}
\includegraphics[scale=0.56]{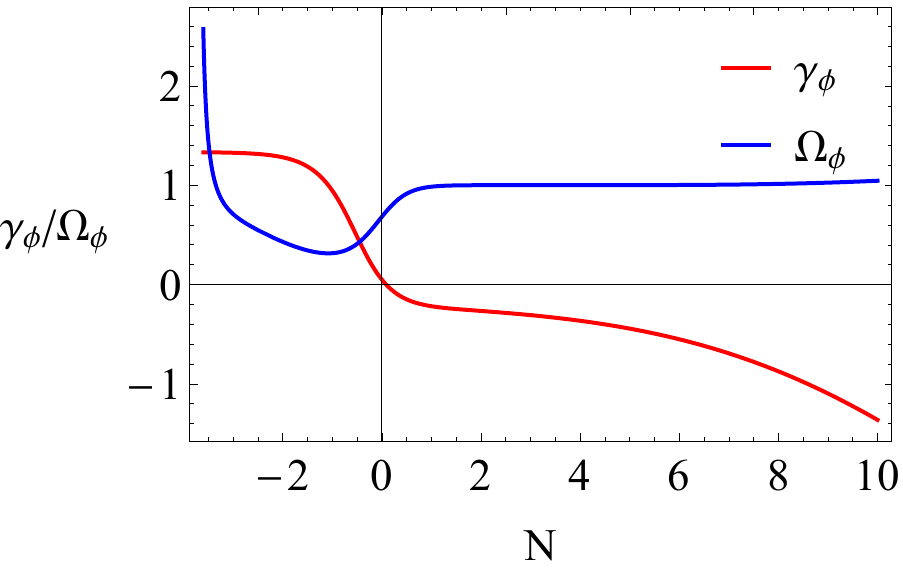}\ \\
\caption{\label{gamma-v7c} The evolutions of $\gamma_\phi$ and $\Omega_\phi$ with N for pure k-essence with dynamical changing $\lambda$ and $\delta$. From left to right, $\Gamma=\tau=0.5,1.4,1.6$.
}}
\end{figure}
\begin{figure}
\center{
\includegraphics[scale=0.6]{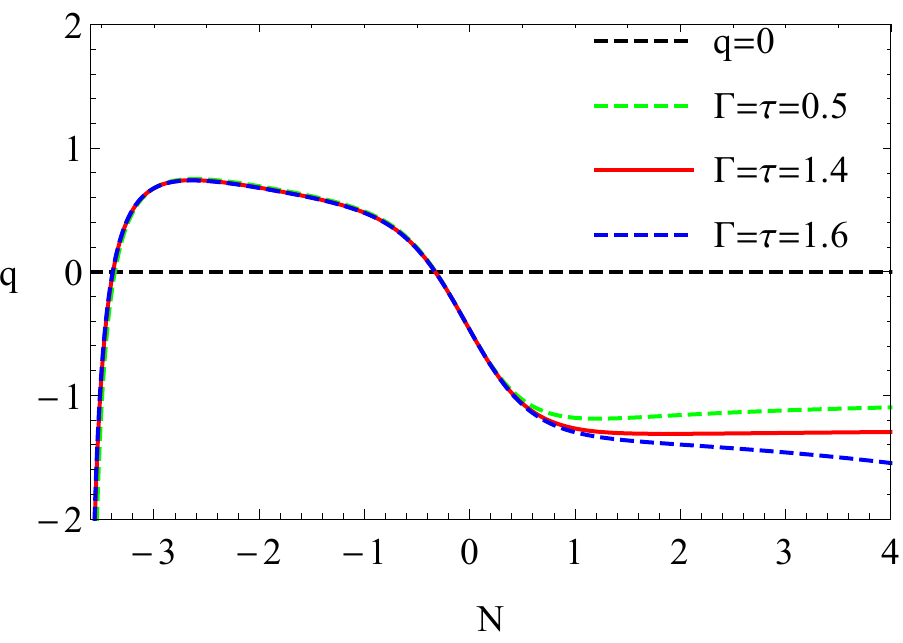}\ \\
\caption{\label{qvsN-v7c} The evolutions of $q$ with N for pure k-essence with dynamical changing $\lambda$ and $\delta$.
}}
\end{figure}

Except the fixed point E, all fixed points are unstable because there is at least one positive eigenvalue for these fixed points. Since there are two negative eigenvalues and three zero eigenvalues for the fixed point E, it is a saddle point. These results for LQC are similar to the ones for the standard FRW universe\cite{LQGRovelli,LQGThiemann,Mulryne:2006cz}.

We want to know the properties of the saddle point E. To this end, we show the evolutions of the system, the parameters $\Omega_\phi$, $\gamma_\phi$ and $q$ with N for different $\Gamma$ and $\tau$ (FIG.~\ref{xqvsN-v7c},~\ref{gamma-v7c} and \ref{qvsN-v7c}). The properties are summarized as what follows.
\begin{itemize}
  \item When $\Gamma$ and $\tau$ is small, the system flows to the fixed point E at the later stage (the first plot in FIG.~\ref{xqvsN-v7c}). Correspondingly, $\Omega_\phi\rightarrow 1$ and $\gamma_\phi\rightarrow 0$, which are the corresponding values of the fixed point E (see TABLE~\ref{tb2-3}).
  \item As $\Gamma$ and $\tau$ increase, some of the variables no longer flow to the fixe point E at the later stage (the second plot in FIG.~\ref{xqvsN-v7c}). Some of the variables even become divergent at the later stage when $\Gamma$ and $\tau$ further increase (the third plot in FIG.~\ref{xqvsN-v7c}). Correspondingly, $\gamma_\phi$ begins to deviate from the value of $\gamma_\phi=0$ for $\Gamma=\tau=1.4$ (the second plot in FIG.~\ref{gamma-v7c}) and even divergent when $\Gamma$ and $\tau$ further increase to the value of $\Gamma=\tau=1.6$ (the third plot in FIG.~\ref{gamma-v7c}).
  \item The deceleration parameters $q$ is different for different $\Gamma$ and $\tau$ at the later stage. In particular, when $\Gamma=\tau=1.6$, $q$ tends to divergent at later stage (FIG.~\ref{qvsN-v7c}).
  \item Again, the LQG effect plays a key role only at the early stage.
\end{itemize}
To sum up, the saddle point E is not an attractor. Whether or not the system flows to this fixed point, it depends on the value of $\Gamma$ and $\tau$.

The most general case is the one with nonzero potential and dynamically changing $\lambda$ and $\delta$. In this case, the system is a $6$-dimensional one. The fixed points and the corresponding eigenvalues are worked out in TABLE \ref{tb2-4}. There are $9$ fixed points for this system. But there are only two stable fixed points (C and D) and two saddle points (E and J) when the parameters satisfy certain conditions. Comparing with the case of the pure k-essence, the system possesses two stable fixed points under certain conditions such that we can model the evolution of the universe.

\begin{table}[htp]
\linespread{1.2}
\centering
\caption{Fixed points for k-essence with nonzero potential and dynamically changing $\lambda$ and $\delta$}
\label{tb2-4}
\setlength{\tabcolsep}{1.0 mm}
 \scriptsize
{\begin{tabular}{c|c|c|c|c|c|c|c|c}
   \hline
    Point    & x & y                     &z  &b                     &$\lambda$                                      &$\delta$ &Eigenvalues&Stability Condition \\
    \hline\hline
    A        & 0                   &$-\frac{\sigma}{12}$   &0  &$\frac{4+\sigma}{4}$  &0                                              &0                      &$-\frac{\sigma}{2}$,$-\frac{\sigma}{4}$,$\frac{\sigma}{4}$,$\sigma$,$-3-\sigma$,$-4-\sigma$,                                                                                                                       &Unstable\\

    \hline
    B        & 0                   &$-\frac{\sigma}{12}$   &0  &$\frac{4+\sigma}{4}$  &$\frac{\sigma}{4(-1+\Gamma)}$                  &0    &$-\frac{\sigma}{4}$,$\frac{\sigma}{4}$,$-\frac{(-3+2\Gamma)\sigma}{4(-1+\Gamma)}$,$-3-\sigma$,$-4-\sigma$                                                                                                        &Unstable\\

    \hline
    C        & 0                   &$-\frac{\sigma}{12}$   &0  &$\frac{4+\sigma}{4}$  &0                                              &$\frac{\sigma}{-5+4\tau}$                      &$-\frac{\sigma}{4}$,$\frac{\sigma}{4}$,$-3-\sigma$,$-4-\sigma$,$-\frac{2(-\sigma+\sigma\tau)}{-5+4\tau}$,$-\frac{-\sigma+\sigma\tau}{-5+4\tau}$                                                                                   &$-3<\sigma<0$, $1<\tau<5/4$ \\

    \hline
    D        & 0                   &$-\frac{\sigma}{12}$   &0  &$\frac{4+\sigma}{4}$  &$\frac{\sigma(-1+\tau)}{(-1+\Gamma)(-5+4\tau)}$&$\frac{\sigma}{-5+4\tau}$  &$\frac{\sigma}{4}$,$\sigma$,$-3-\sigma$,$-4-\sigma$,$\frac{-\sigma+\sigma\tau}{-5+4\tau}$,$-\frac{(2\Gamma-3)(\tau-1)\sigma}{(-1+\Gamma)(-5+4\tau)}$                                        &\tabincell{c}{$\tau>5/4$ or $\tau<1$ \\ $1<\Gamma<3/2$ \\ $-3<\sigma<0$} \\

    \hline
    E        &$-\frac{\sigma}{6}$  &0                      &0  &$\frac{6+\sigma}{6}$  &0                                              &0
    &{0,0,$\sigma$,$\sigma$,$-3-\sigma$,$-6-\sigma$}                                                                                                                                                                       & Saddle point for $-3\leq\sigma\leq 0$ \\

    \hline
    F        &0                    &$\frac{1}{3}$          &0  &0                     &0                                              &0
    &-4,2,1,1,1,$4+\sigma$                                                                                                                                                                                             &Unstable\\

    \hline
    G        &0                    &$\frac{1}{3}$          &0  &0                     &0                                              &$\frac{4}{5-4\tau}$
    &$\frac{4(-1+\tau)}{-5+4\tau}$,$\frac{8(-1+\tau)}{-5+4\tau}$,-4,-1,1,$4+\sigma$                                                                                                                                        &Unstable\\

    \hline
    H        &0                    &$\frac{1}{3}$          &0  &0                     &$\frac{1}{1-\Gamma}$                           &0
    &-4,-1,1,1,$\frac{-3+2\Gamma}{-1+\Gamma}$,$4+\sigma$                                                                                                                                                               &Unstable\\

    \hline
    I        &0                    &$\frac{1}{3}$          &0  &0                     &$-\frac{4(-1+\tau)}{(-1+\Gamma)(-5+4\tau)}$    &$\frac{4}{5-4\tau}$
    &$-\frac{4(-1+\tau)}{-5+4\tau}$,$\frac{4(-3+2\Gamma)(-1+\tau)}{(-1+\Gamma)(-5+4\tau)}$,-4,-1,1,$4+\sigma$                                                                                                              &Unstable\\

    \hline
    J        &-2                    &1                      &0  &0                     &0                                              &0
    &-3,-3,0,0,0,$\sigma$     & Saddle point for $\sigma\leq0$ \\

    \hline
    K        &1&0&0&0&0&0   &${-6,-6,3,0,0,6+\sigma}$                                                      &Unstable\\
    \hline
\end{tabular}}
\end{table}

\section{Conclusion and discussion}\label{sec-con-dis}

In this paper, we have studied the dynamics of a k-essence in LQC. We in particular discuss the stability conditions of the fixed points. Comparing with the standard FRW universe, we need an additional dimensionless variable $z\equiv\rho_t/\rho_c$. Notice that nonzero $z$ relates the LQG effect. Our discussion are divided into two main parts. One is that $\lambda$ and $\delta$ are treated as constant coupling parameters. Another is that $\lambda$ and $\delta$ are dynamically changing variables. For every case, we explore the dynamics of the pure k-essence and k-essence with nonzero potential, respectively. We summarize the main properties of the dynamical system as what follows.
\begin{itemize}
  \item $z$ is zero for all fixed point. It means that the LQG effect is diluted at the later stage of the universe. The evolution picture of the system indicates that LQG effect plays a key role only in the early epoch of the universe.
  \item The fixed points in LQC are basically consistent with that in standard FRW cosmology \cite{Chakraborty:2019swx}. For most of the attractor solutions, the stability conditions are consistent with that for the standard FRW universe. But for some special fixed point, for example, the fixed point C in TABLE \ref{tb2-1}, more tighter constraints are imposed thanks to the LQG effect.
\end{itemize}

In LQC framwork, there are several directions deserving further pursuit.
\begin{itemize}
  \item LQG effect is more evident in the early universe than the current universe. So it is interesting to study the evolution of k-essence in the early universe in LQC framework.
  \item We can explore the k-essence dynamical system in spatially curved FRW universe \cite{Li:2010eua}.
  \item It is definitely interesting to study the dynamical system when an interaction term between k-essence and the fluid is included.
  \item One of important dark energy models, different from the scalar field driving ones, is the so-called Chaplygin gap, which unifies the dark energy and dark matter \cite{Kamenshchik:2001cp,Bento:2004uh,Zhang:2005jj}. Its dynamical behavior has also been studied in \cite{Li:2008uv}. It is interesting to extend such studies to the LQC framework such that we can explore the effect of LQG.
  \item It would be more interesting to study the dark energy evolution in the version of modified Friedmann equation proposed in \cite{Ding:2008tq,Yang:2009fp}. The version of modified Friedmann equation in \cite{Ding:2008tq,Yang:2009fp} reduces to the leading order effective one (Eq. \eqref{Fre-eom-lqc}) if the higher correction term are neglected. The higher correction term may result in a qualitatively different scenario from that of the leading order effective theory \eqref{Fre-eom-lqc}.
\end{itemize}

\begin{acknowledgments}

This work is supported by the Natural Science Foundation of China under
Grant Nos. 11775036, and Fok Ying Tung Education Foundation
under Grant No.171006. Jian-Pin Wu is also supported by Top
Talent Support Program from Yangzhou University.

\end{acknowledgments}

\end{document}